\documentclass[twocolumn]{aastex7}

\renewcommand{\baselinestretch}{1.1}
\usepackage{textcomp}
\usepackage{gensymb}
\usepackage{epstopdf}
\usepackage{amsmath}
\usepackage{graphicx}
\usepackage{hyperref}
\usepackage{multirow}
\usepackage[nameinlink]{cleveref}
\usepackage[multiple]{footmisc}
\crefname{paragraph}{\S}{\S\S} 
\defcitealias{kluter:2025a}{Kl{\"u}ter \& Huston et al.}

\shorttitle{HST Survey of the Galactic Bulge} 
\shortauthors{Terry et al.}

\begin{document}
\title{\textbf{An \textit{HST} Wide Field Survey of the Galactic Bulge: Overview, Strategy, and First Results}}

\author[0000-0002-5029-3257]{Sean K. Terry}
\affiliation{Department of Astronomy, University of Maryland, College Park, MD 20742, USA}
\affiliation{Code 667, NASA Goddard Space Flight Center, Greenbelt, MD 20771, USA}
\email[show]{skterry@umd.edu}

\author[0000-0003-2861-3995]{Jay Anderson}
\affiliation{Space Telescope Science Institute, 3700 San Martin Drive, Baltimore, MD 21218, USA}
\email{jayander@stsci.edu}

\author[0000-0002-5627-5471]{Charles A. Beichman}
\affiliation{NASA Exoplanet Science Institute, IPAC, MS 100-22, Caltech, 1200 E. California Blvd, Pasadena, CA 91125}
\affiliation{Jet Propulsion Laboratory, California Institute of Technology, Pasadena, CA 91109}
\email{charles.a.beichman@jpl.nasa.gov}

\author[0000-0001-8043-8413]{David P. Bennett}
\affiliation{Department of Astronomy, University of Maryland, College Park, MD 20742, USA}
\affiliation{Code 667, NASA Goddard Space Flight Center, Greenbelt, MD 20771, USA}
\email{bennett.moa@gmail.com}

\author{Aparna Bhattacharya}
\affiliation{Department of Astronomy, University of Maryland, College Park, MD 20742, USA}
\affiliation{Code 667, NASA Goddard Space Flight Center, Greenbelt, MD 20771, USA}
\email{abhatta5@umd.edu}

\author[0000-0003-0014-3354]{Jean-Philippe Beaulieu}
\affiliation{Sorbonne Universit\'e, CNRS, Institut d’Astrophysique de Paris, IAP, F-75014 Paris, France}
\affiliation{School of Natural Sciences, University of Tasmania, Private Bag 37 Hobart, Tasmania, 7001, Australia}
\email{beaulieu@iap.fr}

\author[0000-0003-0395-9869]{B. Scott Gaudi}
\affiliation{Department of Astronomy, The Ohio State University, Columbus, OH 43210, USA}
\email{gaudi.1@osu.edu}

\author[0000-0003-1665-5709]{Joel Green}
\affiliation{Space Telescope Science Institute, 3700 San Martin Drive, Baltimore, MD 21218, USA}
\email{jgreen@stsci.edu}

\author[0000-0003-4591-3201]{Macy J. Huston}
\affiliation{Department of Astronomy, University of California Berkeley, Berkeley, CA 94720, USA}
\email{mhuston@berkeley.edu}

\author[0000-0001-9611-0009]{Jessica R. Lu}
\affiliation{Department of Astronomy, University of California Berkeley, Berkeley, CA 94720, USA}
\email{jlu.astro@berkeley.edu}

\author[0000-0003-1581-7825]{Ray A. Lucas}
\affiliation{Space Telescope Science Institute, 3700 San Martin Drive, Baltimore, MD 21218, USA}
\email{lucas@stsci.edu}

\author[0000-0001-5825-4431]{David M. Nataf}
\affiliation{Department of Physics \& Astronomy, University of Iowa, Iowa City, IA 52242, USA}
\email{david-nataf@uiowa.edu}

\author[0000-0001-7506-5640]{Matthew T. Penny}
\affiliation{Department of Physics \& Astronomy, Louisiana State University, Baton Rouge, LA 70802, USA}
\email{penny1@lsu.edu}

\author[0000-0002-1530-4870]{Natalia E. Rektsini}
\affiliation{Sorbonne Universit\'e, CNRS, Institut d’Astrophysique de Paris, IAP, F-75014 Paris, France}
\email{natalia.rec1@gmail.com}

\author[0000-0002-0390-5054]{Carolina Rodriguez Sanchez-Vahamonde}
\affiliation{Space Telescope Science Institute, 3700 San Martin Drive, Baltimore, MD 21218, USA}
\email{crodriguez@stsci.edu}

\author[0000-0002-9881-4760]{Aikaterini Vandorou}
\affiliation{Department of Astronomy, University of Maryland, College Park, MD 20742, USA}
\affiliation{Code 667, NASA Goddard Space Flight Center, Greenbelt, MD 20771, USA}
\email{katievan@umd.edu}

\begin{abstract}

\small \noindent We present an \textit{HST} imaging survey of a $1.1$ deg$^2$ sky area toward the Milky Way Galactic Bulge. This field significantly overlaps with the upcoming \textit{Nancy Grace Roman} Galactic Bulge Time Domain Survey (GBTDS). High angular resolution imaging of this area with \textit{HST} before the start of the \textit{Roman} Galactic Exoplanet Survey (\textit{RGES}) will greatly strengthen \textit{Roman}'s ability to characterize detected exoplanet systems, as well as provide a rich and wide-field archive for use as a Legacy dataset toward the Galactic Bulge for the broader community. We conduct coordinated-parallel imaging with both wide-field cameras on \textit{HST}, Wide-field Camera 3 (WFC3) and Advanced Camera for Surveys (ACS), utilizing the $F606W$ and $F814W$ passbands. Approximately 70\% of the survey was conducted during \textit{HST} Cycle 32, with the remaining 30\% conducted during Cycle 33. In this paper, the first in a series, we give a general overview of the program and the observing strategy, and present early results. This campaign secures \textit{HST}'s lasting impact on the high-precision study of stellar populations, dynamics, exoplanet systems, interstellar extinction, metallicities, cluster associations, and more toward the center of our Galaxy.
\\
\\
\textit{Subject headings}: Galaxy: Bulge \textemdash \, Galaxy: stellar content \textemdash \, stars: low-mass \textemdash \, gravitational lensing: micro, planetary systems \textemdash \, Roman\\
\end{abstract}


\section{Introduction} \label{sec:intro}
\indent The central Galactic Bulge of the Milky Way is an environment rich in stars with a broad range of chemical compositions, kinematics, and ages. Given the nearby nature of the bulge, several observatories can study the resolved stellar populations across the region \citep{Holtzman:1998a, skrutskie:2006, brown:2009a, brown:2010a, gonzalez:2012a, nataf:2013a, hambleton:2023a}. Understanding the properties of this population is fundamental to our understanding of Galaxy formation, evolution, dynamics, and more. One crucial property of this stellar population that has recently been debated is the true age distribution of these stars. Prior studies that derive ages from bulge color-magnitude diagrams (CMDs) suggest that the bulge has a uniformly old age (${\sim}\, 10$Gyr) \citep{ortolani:1995a, Clarkson:2008a, Gennaro:2015a}. In contrast, recent spectroscopic analysis of this population suggests at least 15\% may be metal-rich dwarf stars younger than $5-8$ Gyr \citep{bensby:2017a}. Another critical metric for understanding the formation and evolution of our Galactic bulge is the total mass budget in the bulge. While historical attempts to measure the mass-to-light (M/L) ratio have yielded bulge masses that differ by a factor of $2-3$ \citep{sellwood:1988a, kent:1992a}, more recent studies have converged on a consistent estimate of the bulge dynamical mass. \cite{portail:2015a} and \cite{portail:2016a} used a dynamical model of the bulge with stellar kinematics from BRAVA \citep{Rich:2007a, kunder:2012a} and 3D surface brightnesses from \cite{wegg:2013a} to estimate a bulge dynamical mass of $1.8\pm0.07\times10^{10}\, M_{\odot}$.\\
\indent There have been many additional studies of the galactic bulge region including; investigations of the galactic bar structure \citep{stanek:1994a, stanek:1997a, pietrukowicz:2015a, simion:2017a}, interstellar extinction and reddening curves toward the bulge \citep{nataf:2010a, nataf:2013a, gonzalez:2012a, Surot:2020a}, the initial mass function, star formation rate and history of the bulge \citep{Calamida:2015a, haywood:2016a, bernard:2018a}, and surface densities and stellar proper motions \citep{Clarkson:2008a, terry:2020a}.\\
\indent Lines of sight toward the Galactic bulge are also excellent targets for gravitational microlensing surveys \citep{udalski:1993a, bond:2001a, kim:2016a}, which can be used to probe the mass and kinematic distributions of stars in the Galactic bulge and foreground disk \citep{moniez:2010a}, as well as to search for exoplanets \citep{mao:1991a, gould:1992a, gaudi:2012a}. Microlensing requires an extremely precise (${\sim}1$ mas) chance alignment of a foreground mass (the lens) with a background star (the source) on the sky. As a result, both the optical depth to microlensing and the microlensing event rate scale with the square of the stellar surface density, and a productive microlensing survey therefore requires very dense stellar fields to achieve a suitable event rate and optical depth \citep{Mao:2008a}. Among nearby resolved stellar populations, the optical depth and event rate are largest toward the Galactic bulge \citep{paczynski:1991a, kiraga:1994a}. Owing to the competing effects of extinction and stellar surface density with galactic latitude, optical microlensing surveys find that the event rate peaks a few degrees above and below the Galactic plane \citep{sumi:2003a, mroz:2019a}.\\
\indent One of the deepest observational campaigns ever conducted toward the Galactic bulge will be the Galactic Bulge Time Domain Survey (GBTDS), one of several core community surveys to be carried out by the upcoming Nancy Grace Roman Space Telescope (formerly WFIRST; \cite{Spergel:2015a}). In its primary survey area, the GBTDS will monitor five fields of a total of approximately 1.4 square degrees using the Wide Field Instrument (WFI), with a cadence of approximately 12 minutes in the wide 1$-$2 $\mu$m F146 filter. The survey will span six seasons of 72 days each, for a total duration of 438 days. Over this period, \textit{Roman} will achieve relative photometric precision of ${\lesssim}$1\% and relative astrometric precision of ${\lesssim}$1 mas per epoch over approximately 50,000 epochs for roughly $10^8$ stars. A sixth field centered on the Galactic center will also be monitored concurrently with the five main survey fields. As part of this bulge survey, the \textit{Roman} Galactic Exoplanet Survey (\textit{RGES}) will be the first dedicated space-based gravitational microlensing survey and is expected to detect over 30,000 microlensing events and over 1,400 bound exoplanets during its five-year survey \citep{penny19, terry:2025a, saggese:2025a, zohrabi:inprep}. The survey is also expected to discover several hundred free-floating planets (FFPs) and/or planets on very wide orbits around host stars \citep{johnson:2020a, sumi:2023a, lastovka:2025a}. Detailed knowledge of the stellar environment and extinction in the expected bulge fields will be needed prior to the start of the \textit{RGES} in order to optimize the scientific output of the mission. \\
\indent Beyond exoplanet detection, the GBTDS will enable a wide range of additional science \citep{gaudi:2019a}, including the detection of roughly 100,000 transiting planets \citep{montet:2017a, wilson:2023a}, asteroseismic oscillations in approximately 290,000 giant stars \citep{gould:2014a, weiss:2025a}, and monitor about 5,000 trans-Neptunian objects \citep{gould:2014b}. The survey will also measure parallaxes and proper motions for roughly six million bulge and disk stars \citep{sanderson:2019a}, enable the detection of millions of variable stars, and generally support many additional investigations.\\
\indent Given the substantial resources devoted to the GBTDS, it is important to leverage complementary datasets to maximize its scientific return. A unique aspect of the GBTDS, relative to other bulge surveys, is its combination of large areal coverage and high angular resolution. In particular, \textit{Roman} will resolve hundreds of millions of main-sequence stars \citep{penny:2019a}. Images at other wavelengths, or at times significantly before or after the survey, that also resolve and detect a significant fraction of these stars, will provide enormous additional information about the stellar populations and extinction. This information can be used to optimize the scientific output of the survey and to aid in the interpretation of its results. \\
\indent At the surface densities of main-sequence stars in the GBTDS field of view, and with \textit{Roman}’s angular resolution of approximately 0.1 arcseconds, stars that are unresolved or partially resolved (that is, blended) are likely to be physically associated systems. In particular, a star blended with a microlensing source is mostly likely to be a companion to the lens, the source, or the lens host itself, with a smaller chance of being an unrelated ambient field star \citep{bhattacharya:2017a}. Given typical bulge proper motions of $2–10$ mas yr$^{-1}$ and a time baseline of roughly five years, it will be possible to measure proper motions for nearly all stars with single-epoch astrometric uncertainties better than about 1 mas. Consequently, the proper motions and fluxes of many stars in the GBTDS, including a large fraction of lens hosts, can be measured. As discussed in more detail below, this enables estimates of host and planet masses \citep{bennett:2007a}, which are critical for comparisons with planet demographics derived from other detection methods and for testing planet formation theories. These characterization techniques are substantially enhanced when observations at comparable resolution and sensitivity are available in additional filters and/or over longer time baselines than the nominal five-year \textit{Roman} mission. \\
\indent In terms of angular resolution and sensitivity, the \textit{Hubble Space Telescope (HST)} is exceptionally well matched to \textit{Roman}. In addition, its imaging instruments provide access to bluer wavelengths that are complementary to \textit{Roman}’s near-infrared coverage. \textit{HST} has been widely used to image the Galactic bulge along multiple sight lines to study stellar populations, ages, kinematics, and star formation history. Using multi-epoch \textit{HST} observations, stellar proper motions for disk and bulge population stars have been studied by \cite{Kozlowski:2006a} and \cite{Clarkson:2008a}. Age and metallicity estimates of globular clusters have been performed by \cite{Milone:2012a}, \cite{Lagioia:2014a}, \cite{Calamida:2014a}, and \cite{Baldwin:2016a}, building on the observations of \cite{brown:2009a, brown:2010a}. A broad study of the star formation rate and initial mass function was conducted using predominantly OGLE29 field images at $(\ell, b)$ = (-6.753, -4.720; \cite{Gennaro:2015a}). The star formation histories of the Stanek and SWEEPS fields, derived from Wide Field Camera 3 (WFC3) photometry, were studied by \cite{bernard:2018a}, who found them to be remarkably similar. \textit{HST} has also been used extensively to characterize host stars of planets detected in ground-based microlensing surveys \citep{dong:2009a, bennett:2015a, bhattacharya:2018a, bhattacharya:2023a, terry:2024a}. \\
\indent Because of the relatively small field of view of \textit{HST} imaging cameras, these studies have primarily been pencil-beam surveys of targeted bulge regions and therefore cover only a small fraction of the total bulge area. There is, however, strong motivation to devote a large number of \textit{HST} orbits to a blind survey covering a significant fraction of the GBTDS main survey area prior to the launch of \textit{Roman}. Although \textit{HST} continues to operate and produce cutting-edge science, there is no guarantee that it will remain operational for another decade, making it prudent to undertake such a survey as soon as possible. Moreover, early, untargeted imaging of a large fraction of the field enables the measurement of proper motions for the host stars of planetary microlensing events that will occur in the future during the \textit{Roman} GBTDS. \\
\indent This motivation underlies the large \textit{HST} program described here (GO-17776; \cite{terry:2024prop}), a wide-area survey covering a substantial fraction (${\sim}$70\%) of the GBTDS footprint using full-frame images from two cameras operated in parallel. While the scope of the project and its data products is broad, the program is largely designed to support the \textit{Roman} GBTDS mission. This paper is organized as follows. We discuss the overall strategy and some specific objectives of the \textit{HST} survey in Section \ref{sec:goals}. In Section \ref{sec:obs_reductions} we present the \textit{HST} field selections and observing strategy, and we compare this to the expected \textit{Roman} strategy. In Section \ref{sec:first-look} we describe the anticipated data products from the program, and we show preliminary results from a few early fields observed by the survey. We conclude the paper in Section \ref{sec:conclusion} with a discussion of the initial results and implications for the \textit{Roman} survey.


\section{Goals of the \textit{HST} Bulge Survey} \label{sec:goals}

\subsection{Preparing for \textit{Roman}} \label{subsec:roman-prep}
NASA's next flagship mission, the \textit{Nancy Grace Roman Space Telescope}, is scheduled to launch no later than Spring 2027. One of the core community surveys to be conducted in the first five years of \textit{Roman} is the GBTDS, which will utilize the wide-field imager (WFI) to monitor ${\sim}2$ square degrees toward the center of the galaxy. One of the primary science drivers of the GBTDS is the discovery and characterization of exoplanets via the microlensing technique. However there are expected to be many other general astrophysics (GA) related studies conducted on the deep, time-series \textit{Roman} data that the GBTDS provides. \\
\indent The \textit{Roman} Observations Time Allocation Committee (ROTAC)\footnote[2]{\url{https://roman.gsfc.nasa.gov/science/ccs/ROTAC-Report-20250424-v1.pdf}} has recommended five contiguous \textit{Roman} WFI fields placed in the southern Galactic Bulge region, and one \textit{Roman} WFI field placed on the Galactic Center (GC). Table \ref{tab:survey-params} shows the ROTAC-recommended survey parameters for the GBTDS as compared to the survey parameters for our \textit{HST} survey. Although there is still a small possibility of survey modification (e.g. final filter selection for the GBTDS), the survey is now essentially defined. As discussed in the following subsections, we are motivated to study these GBTDS fields in a precursor context to gain useful insight on the bulge stellar Luminosity Function (LF), interstellar extinction, and reddening (which affects filter selection and exposure times), astrometric baselines for future proper motion studies with \textit{Roman}, and a potential large-scale input catalog of ${\sim}$dozens of millions of stars in these fields with high precision photometric and astrometric measurements delivered by \textit{HST}.\\

\begin{deluxetable*}{lcc}[!htp]
\deluxetablecaption{\textit{HST} Survey Parameters Compared to the GBTDS\label{tab:survey-params}}
\tablecolumns{3}
\setlength{\tabcolsep}{34pt}
\tablewidth{\columnwidth}
\tablehead{
\colhead{\hspace{-1.8cm}Parameter} & \colhead{\textit{HST}} & \colhead{\textit{Roman GBTDS}}
}
\startdata
Area [deg$^2$] & 1.12 & 1.41 (five-fields)\\
Seasons & 1 & 6\\
Fields & 354 & 6\\
Epoch of Obs. & $2025-2026$ & $(2027-2031)^a$\\
Cadence [min] & Single visit & $12.1^b$\\
Primary Filters & $F606W, F814W$ & $F087, F146, F213$ \\
Total exposures & 4 per field & ${\sim}$8,400 per field\\
Photometric precision & 0.04 mag $@\, F814W{\sim}19.8$ & (0.02 mag $@\, F146\,{\sim}\,21.2)^c$\\
Astrometric precision & 3 mas $@\, F814W{\sim}19.1$ & (1 mas $@\, F146\,{\sim}\,21.9)^d$\\
\enddata
\tablenotetext{}{\footnotesize{$^a$ Based on the notional survey timeline, subject to change.\\
$^b$ High-cadence GBTDS seasons.\\
$^c$ \cite{wilson:2023a, terry:2023b} \\
$^d$ \cite{sanderson:2019a, terry:2023b}}}
\end{deluxetable*}

\begin{figure*}[!htb]
\includegraphics[width=0.8\linewidth]{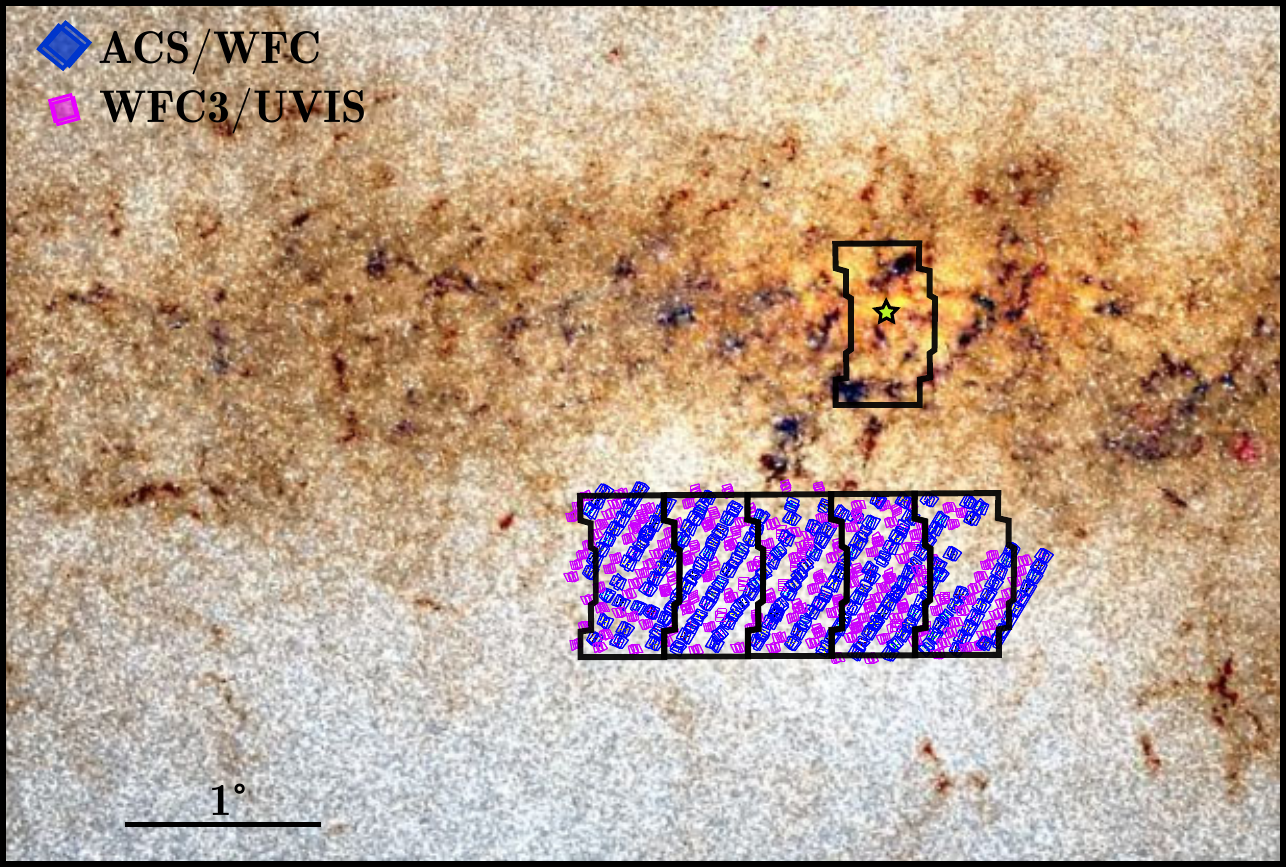}
\centering
\caption{The central Galactic bulge sky region. We show the \textit{HST} bulge survey footprint, where blue FoVs are ACS/WFC and purple FoVs are WFC3/UVIS. Underlaid is the 2 Micron All-Sky Survey (2MASS) \citep{skrutskie:2006}. Black outlines show the \textit{Roman} GBTDS observing fields. The \textit{HST} footprint covers approximately 1.1 deg$^2$ (see Table \ref{tab:survey-params}).} \label{fig:mosaic}
\end{figure*}

\begin{figure}[!htb]
\includegraphics[width=\linewidth]{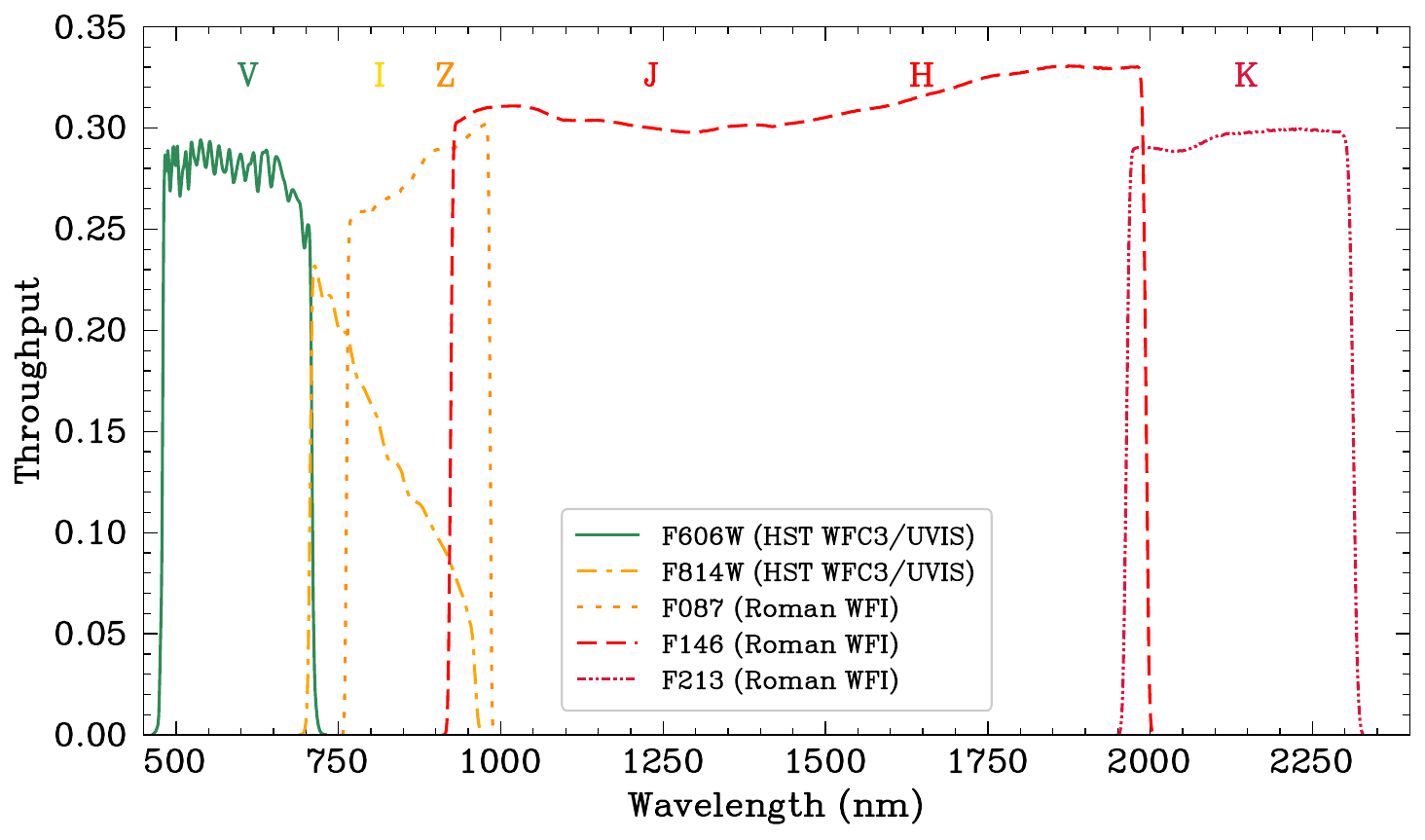}
\centering
\caption{Filter throughputs for the passbands used in the \textit{HST} bulge survey ($F606W, F814W$ on WFC3/UVIS) as well as the three primary filters that will be used for the \textit{Roman} GBTDS ($F087, F146, F213$).} \label{fig:filters}
\end{figure}

\subsection{Microlensing Lens Flux Measurements} \label{sec:precursor-lens-detections}
One of the primary level-1 science requirements for the \textit{RGES} is to determine the masses of, and distances to, host stars of 40\% of the detected microlensing planets with a precision of 20\% or better. One of the primary motivations of our \textit{HST} survey is to increase the fraction of detected planetary systems with accurate mass and distance estimates by enabling precursor detection of lenses and sources several years before they align to cause a microlensing signal detected by \textit{Roman} during the GBTDS. \\
\indent From a microlensing light curve alone, it is possible to obtain a “full solution” for the lens system, meaning a measurement of the mass and distance to the lens, by measuring two higher order effects on the light curve that are detectable in a subset of events. The first is due to the finite size of the source, which “smooths” over regions of sharp changes in the magnification for a point source.  The magnitude of the smoothing depends on the angular size of the star $\theta_*$ in units of the angular Einstein ring radius of the lens $\rho = \theta_*/\theta_{\rm E}$, where the angular Einstein ring radius sets the scale of the lensing system and is given by,
\begin{equation}  
\theta_E \equiv \sqrt{\kappa M_L \pi_{\rm rel}}
\end{equation}
\noindent where $M_L$ is the mass of the lens, $\kappa = 4G/(c^2~{\rm AU})\simeq 8.144~{\rm mas}M_\odot^{-1}$, $\pi_{\rm rel} \equiv \pi_L - \pi_S$ is the relative lens-source parallax, $\pi_L={\rm AU}/D_L$ and $\pi_S={\rm AU}/D_S$ are the parallaxes of the lens and source, and $D_L$ and $D_S$ are the distances to the lens and source.  A measurement of $\theta_{\rm E}$  allows one to derive a relation between the lens mass and distance given an estimate of the distance to the source,

\begin{equation} \label{eq:md_thetaE}
    M_{L} = \frac{c^2}{4G}\theta_{E}^{2}\frac{D_{S}D_{L}}{D_{S}-D_{L}},
\end{equation}

\noindent The second is due to the non-uniform relative proper motion between the lens and source due to the acceleration of the observatory around the sun, which allows one to measure the microlensing parallax $\vec{\pi}_E$, which has a magnitude
\begin{equation}
    \pi_E = \frac{\pi_{\rm rel}}{\theta_E},
\end{equation}

\noindent and direction equal to that of the relative lens-source proper motion $\vec{\mu}_{\rm rel}$. It is useful to note that the magnitude of the lens-source proper motion is related to the Einstein timescale of the event $t_{\rm E}$, another observable quantity of microlensing light curves, by $\mu_{\rm rel} = \theta_E/t_E$.  Note that $\vec{\pi}_E$ can also be measured from differences in the light curves seen by two non-cospatial observatories, such as Earth-L2 \citep{wyrzykowski:2020a} or multiple locations on the Earth \citep{gould:2009a}. A measurement of $\pi_E$ allows one to derive another relation between the lens mass and distance given an estimate of the distance to the source,

\begin{equation} \label{eq:md_piE}
    M_{L} = \frac{c^2}{4G}\frac{\textrm{AU}}{{\pi_E}^2}\frac{D_{S}-D_{L}}{D_{S}D_{L}},
\end{equation}

\noindent where $M_{L}$ is the lens mass, and $G$ and $c$ are the gravitational constant and speed of light. $D_{L}$ and $D_{S}$ are the distance to the lens and source, respectively. Combining these two equations yields the mass of the lens with no dependence on the lens or source distance:

\begin{equation} \label{eq:md_thetaE_piE}
    M_{L} = \frac{c^2 \theta_E \textrm{AU}}{4G \pi_E} = \frac{\theta_E}{(8.1439\, \textrm{mas})\pi_E}M_{\odot}.
\end{equation}

\noindent One can calculate the mass of the lens assuming a clear measurement of both higher-order effects, but this is a relatively rare situation. Additionally, if a measurement of flux coming directly from the lens host can be made, then empirical mass-luminosity relations (i.e. \cite{henry:1993a, delfosse:2000a}) can be used to obtain a third mass-distance relation. This relation is given by the following:

\begin{equation}\label{eq:md-3}
    {\cal F}_\lambda = \frac{L_\lambda(M_L)}{4\pi D_L^2} \ 
\end{equation}

\noindent Here $L_\lambda(M_L)$ is the relationship between the mass of a star and its specific luminosity, i.e., its luminosity in a given filter or passband. It is worth noting that this relationship depends not only on the specific passband, but in principle also depends on other properties of the star such as its metallicity, surface gravity, age, etc., which are generally unknown. It is easiest to independently measure the source and lens stars if a sufficient amount of time has passed after (or before) the microlensing event. Some of the earliest examples of direct lens star detections via follow-up \textit{HST} imaging include targets toward the LMC \citep{alcock:2001a} and Galactic bulge \citep{kozlowski:2007a}. The multi-passband \textit{HST} observations from our survey will significantly enhance the efforts to obtain this third mass-distance relation for microlensing hosts and planetary companions that will be detected in future \textit{Roman} data (see Section \ref{subsec:euclid-roman}).\\
\indent The direct lens flux measurement is beneficial in several aspects beyond just the mass-distance relation. If the lens and source stars can be independently measured, then their separations on the sky can be calculated which leads to a direct measurement of their relative proper motions ($\vec{\mu}_{\textrm{rel}}$). When both components of $\vec{\mu}_{\textrm{rel}}$ (East and North) are measured well, a tighter constraint can be placed on the possible microlensing parallax values, particularly $\pi_{\textrm{E,N}}$, the North component of the microlensing parallax vector. For typical microlensing events observed toward the Galactic bulge, the North direction of the microlensing parallax is only weakly constrained during the light curve modeling because this direction is approximately perpendicular to the orbital acceleration of the observer. This will also be the case with modeling microlensing events from \textit{RGES} at L2. \\
\indent Ideally, an \textit{HST} precursor imaging campaign of the GBTDS fields would have been conducted at least five or ten years ago when the lens and source star separations were larger than they are at present. Nevertheless, the observations presented here, taken only a year or two before the first GBTDS season, are still enormously beneficial. In Section \ref{subsec:euclid-roman} we quantify the relation between lens detections and the time baseline ($\Delta t$) between microlensing peak and high-resolution followup (or precursor) imaging. The exquisite stability of \textit{HST}'s Point Spread Function (PSF) allows for highly blended lens-source pairs at very small separations (e.g. ${\sim}\,12$ mas) to be measured (see \citealt{bhattacharya:2017a}). We expect the elongation of the blended stellar images to be detectable in \textit{HST} for many of the microlensing events that \textit{Roman} will discover during its survey. Further, the color-dependent centroid shift method \citep{bennett:2007a} can be employed to independently measure the source star and lens star fluxes. These \textit{HST} color measurements can also help disentangle the stellar types for highly blended source, lens, and nearby field stars along the galactic bulge sight lines. Lastly, these precursor \textit{HST} observations will image sources and lenses in a standard visible passband ($F606W$) which is largely unreachable by \textit{Roman}'s near-IR detectors (see Figure \ref{fig:filters}). In the \textit{Roman} era, new color-surface brightness relations will need to be derived for \textit{Roman} passbands to ensure accurate estimates of the source star angular sizes ($\theta_*$) are made. Accurately measuring this quantity allows for an estimate of the angular Einstein radius $\theta_E$. These quantities are related through the source radius crossing time, which is the time it takes the lens-source relative motion to traverse one source star radii. The expression is $t_*=\rho\, t_E$, where $\rho=\theta_*/\theta_E$. This means that $t_*$ effectively sets the duration of finite source effects in a light curve, and our ability to precisely measure this quantity clearly depends how well one can determine the source size in physical units (typically $\mu$as). Well-understood color-surface brightness relations have been used in past stellar population studies as well as microlensing analyses from the ground \citep{kervella:2004a, boyajian:2014a, adams:2018a}. The \textit{HST} $V$ and $I-$band measurements of \textit{Roman}-detected microlensing events can serve to help properly calibrate the near-IR color-surface brightness relations during the \textit{RGES}.

\begin{figure*}[!htb]
\includegraphics[width=0.7\linewidth]{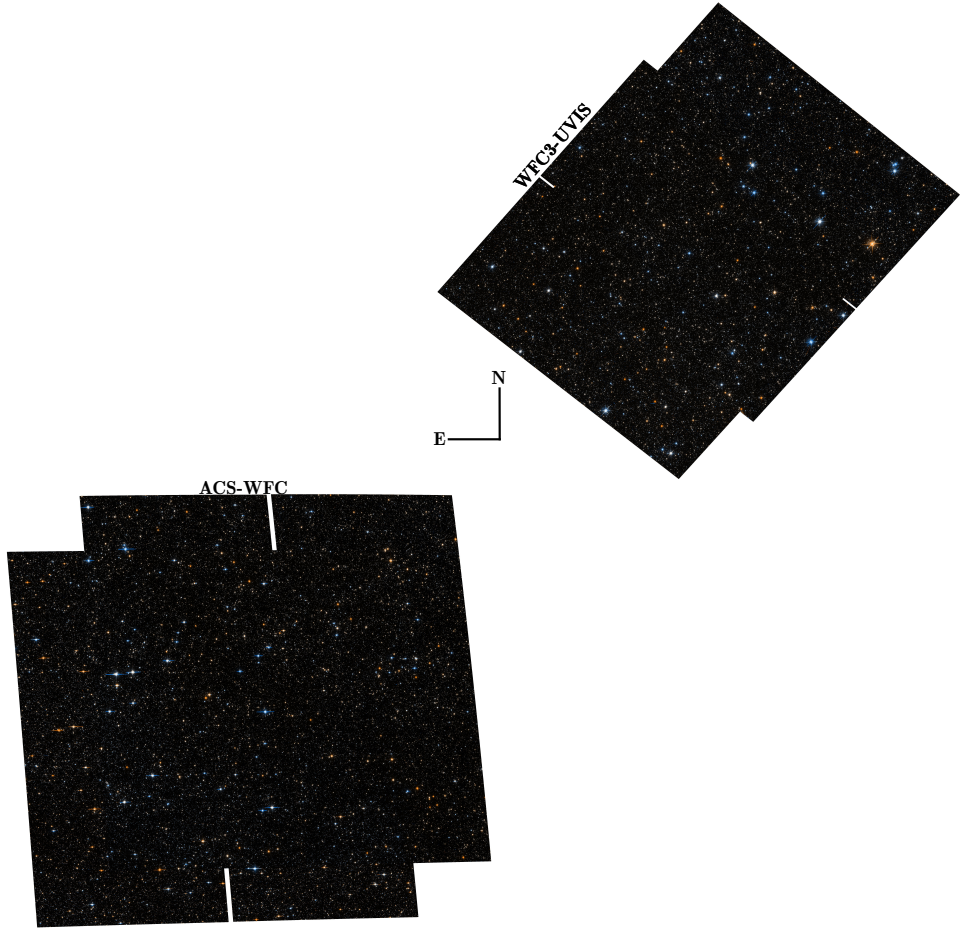}
\centering
\caption{Color composite of one parallel field (HD70) imaged during the \textit{HST} survey. The position angle of WFC3/UVIS is rotated approximately 45 degrees relative to ACS/WFC, and the center of the focal planes are separated by ${\sim}$6 arcminutes on the sky. ACS/WFC has a total FoV that is ${\sim}$40\% larger than WFC3/UVIS.}
\label{fig:hd70_composite}
\end{figure*}

\begin{figure*}[!htb]
\includegraphics[width=0.6\linewidth]{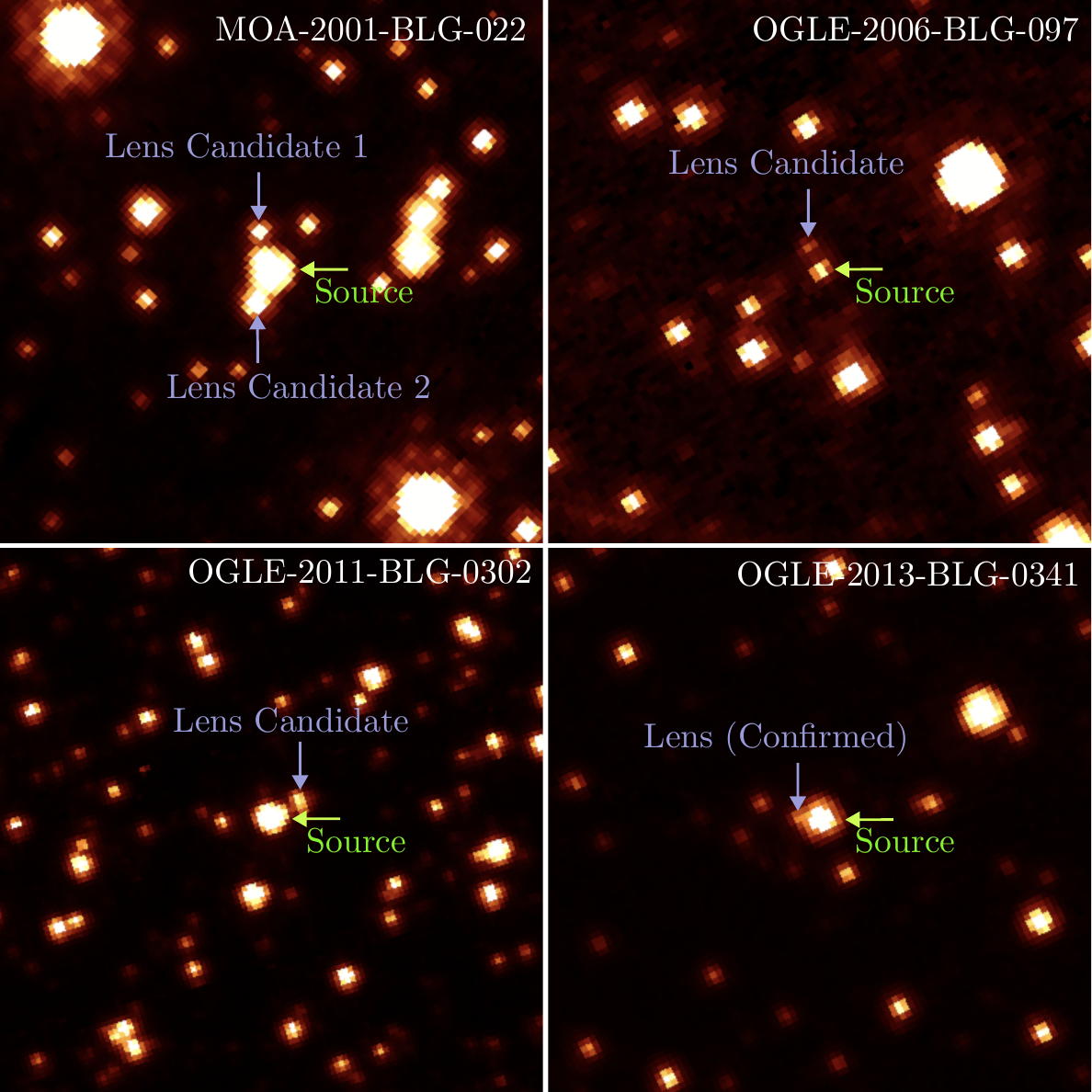}
\centering
\caption{\textit{HST} reference images from four fields that contain historical microlensing events. Source stars are labeled with yellow arrows, lens candidates are labeled with blue arrows. One candidate, OGLE-2013-BLG-0341L (\textit{bottom-right panel}), has been confirmed as the true lens via Keck adaptive optics (AO) and \textit{Euclid} data \citep{rektsini:inprep}.} \label{fig:quad_events}
\end{figure*}

\subsection{Historical Microlensing Events in the \textit{Roman/HST} Footprint} \label{subsec:historical_events}
There is a long history of microlensing campaigns being conducted toward these highly-crowded galactic bulge fields. The Optical Gravitational Lensing Experiment (OGLE; \cite{udalski:1993a, udalski:2015a}) and Microlensing Observations in Astrophysics (MOA; \cite{bond:2001a, sumi:2003a}) projects have been monitoring the galactic bulge region for three decades. More recently, the Korea Microlensing Telescope Network (KMTNet; \cite{kim:2016a}) has been monitoring the bulge fields with a network of three ground-based telescopes located in Australia, South Africa, and South America. This network has enabled high-cadence, nearly continuous monitoring of the stars toward the bulge. In total, approximately 40,000 microlensing events have been detected by these surveys, of which approximately 270 have been found to have exoplanet signals in their light curves.\\
\indent Since this part of the sky is ideal for microlensing surveys, the \textit{Roman} GBTDS and \textit{HST} precursor survey both have complete overlap of the sky area searched by these ground-based campaigns. Naturally, this means that there are many thousands of historical microlensing events originally detected from the ground that lie within our \textit{HST} precursor survey. In some cases, over 20 years have passed since the original microlensing event was detected from the ground. This means the sources and lenses for these historical events have a high likelihood of being resolved in the new \textit{HST} high-resolution data \citep{bennett:2024a, terry:2024a, rektsini:inprep, terry:inprep}. As mentioned in Section \ref{sec:precursor-lens-detections}, a measurement of the lens flux is crucial in determining several physical properties of the lens system (e.g. host mass, planet mass, distance from Earth). \\
\indent One caveat that may complicate the analysis for some of the significantly older historical events is that `too much' time has passed since the microlensing event, such that the identification of the lens star is ambiguous and possibly confused with other nearby unrelated stars in the field. \cite{Holtzman:1998a}, \cite{Calamida:2015a}, and \cite{terry:2020a} estimate that there are $10^3-10^4$ stars arcmin$^{-2}$ with magnitudes fainter than $M_I > 12$ toward various galactic bulge fields (Baade's Window, SWEEPS, Stanek). The \textit{HST/Roman} GBTDS fields are located in a sky area with ${\sim}\,2\times$ higher surface densities than these previously studied sight lines, which increases the probability of confusion when attempting to identify the lens stars for historical events. If an estimate of the lens-source relative proper motion ($\mu_{\textrm{rel,G}}$) can be made using the light curve data, it is only the magnitude of the relative motion that is estimated, not the two-dimensional direction. For the confused events in this regime, an additional epoch of high-resolution imaging (e.g. from \textit{Roman} itself) will allow us to trace the 2D relative proper motions ($\mu_{\textrm{rel,N}}$, $\mu_{\textrm{rel,E}}$) back in time and determine which of the lens candidates was located at the approximate position of the source star during the time of peak magnification.\\
\indent The top-left panel of Figure \ref{fig:quad_events} shows one such example of this type of confusion. The brighter source star is at center frame indicated by the yellow arrow, with two possible nearby lens candidates indicated with blue arrows. The two candidates have a similar on-sky separation from the source star, which leads to the ambiguity discussed above. This precursor \textit{HST} observation can be coupled with \textit{Roman} observations in GBTDS seasons one and two to reliably determine which candidate is the true lens star.

\subsection{Input Catalog for the \textit{Roman} GBTDS} \label{subsubsec:input_cat}
The GBTDS is expected to discover more than one thousand microlensing planets \citep{penny:2019a}, hundreds of FFPs \citep{johnson:2020a, sumi:2023a}, and possibly one hundred thousand transiting planets \citep{montet:2017a, wilson:2023a}. The knowledge and characterizability of the bound planets will strongly depend on our ability to understand their host star properties. Prior transiting exoplanet missions like \textit{Kepler} and \textit{TESS} owe much of their success to the \textit{Kepler} Input Catalog (KIC) \citep{brown:2011a} and \textit{TESS} Input Catalog (TIC) \citep{stassun:2019a}. These input catalogs are able to leverage photometry in multiple optical bands to constrain host star properties like effective temperatures, surface gravities, radii, and masses. All of this information is essential for deriving reliable planet occurrence rate estimates, which have been the most impactful results coming out of these transiting planet surveys. \\
\indent Ultimately, \textit{Roman} itself will produce one of the deepest stellar catalogs ever generated by the end of the nominal GBTDS. The table of ${\sim}25$ million stars produced by our \textit{HST} survey can be used as an input catalog for \textit{Roman} during the first season of the GBTDS expected in 2027. As was done for \textit{TESS}, a maximal amount of multi-wavelength photometry can be included in such an input catalog \citep{stassun:2019a}, including measurements of stellar properties from other photometric surveys like \textit{Gaia} \citep{vallenari:2023a}, VVV \citep{saito:2012a,minniti:2010a}, DECaPS \citep{saydjari:2023a}, 2MASS \citep{skrutskie:2006a}, WISE \citep{wright:2010a}, and more. Lastly, one can use existing stellar characterization tools such as \texttt{isochrones} \citep{morton:2015a}, \texttt{isoclassify} \citep{huber:2017a}, or tools specifically developed for \textit{Roman} such as \texttt{brutus} \citep{speagle:2025a} to combine all information to determine stellar luminosities, radii, and angular diameters, and place constraints on stellar masses, ages, and line-of-sight extinction. Some of these parameters will be more difficult to constrain (masses, ages) than others (angular diameters, extinction). Ultimately \textit{Roman} itself will be the `gold standard' catalog, and will likely include stellar distances for many of the stars via geometric parallax measurements. We note that the development of a fully complete, multi-observatory input catalog is beyond the scope of our current work. We intend to publish our \textit{HST} point source catalog with color measurements, astrometric positions, and quality-of-fit metrics in Paper 2 of this series. All published catalogs from our survey will be delivered in machine-readable formats for use by the broader community, and will be easy to query via a simple \texttt{python}-based tool \citep{terry:inprep}.

\subsection{Luminosity Functions and Extinction Maps} \label{subsec:extinction}
Luminosity Functions (LFs) are essential in understanding many different characteristics of stellar populations, particularly the stellar surface density, initial mass function, and mass-to-light (M/L) ratio. The mass function for the Galactic bulge was first measured by \cite{Holtzman:1998a} and was based on \textit{HST}-WFPC2 observations of Baade's Window [$l,b$] = ($1.1\degree, -3.8\degree$). They derived a LF down to $F814W\,{\sim}\,24$ which corresponds to a mass of ${\sim}0.3M_{\odot}$. Further, the study found that the Galactic bulge Initial mass function has roughly a power-law slope of $\alpha=-2.2$ at high-mass, and begins to flatten at ${\sim}0.7M_{\odot}$. Subsequent \textit{HST} studies have been conducted in the Baade's, SWEEPS, and Stanek fields \citep{zoccali:2000a, Calamida:2014a, terry:2020a} to derive LFs toward nearby sight-lines, and to attempt to disentangle the foreground disk star population from the background bulge population. We note these previous \textit{HST} studies give mass functions and/or LFs in fields that do not directly overlap with the expected GBTDS fields.\\
\indent The interstellar extinction toward the central region of the Galaxy is highly variable and can be as large as $A_V > 20$ at the lowest latitudes. There has been significant advancement in our understanding of the extinction and reddening toward the galactic bulge in the past several decades, particularly via the study of RR Lyrae and red clump (RC) stars using wide-field ground-based telescopes \citep{gonzalez:2012a, nataf:2013a, Surot:2020a} as well as \textit{HST} \citep{revnivtsev:2010a}. Until recently, a major issue plaguing the interpretation of extinction and reddening toward the bulge was the fact that the extinction curve is variable on relatively small spatial scales across the region. It was not until \cite{nataf:2016a} combined the $E(V-I)$ reddening maps from OGLE-III \citep{udalski:1993a} with the $E(J-K)$ reddening maps from VVV \citep{gonzalez:2012a} that a robust measurement of $R_V$ and its spatial variation was made across the bulge region. \cite{nataf:2016a} also included measurements of $E(I-J)$, which provided further constraints on the wavelength dependence of the extinction curve.\\
\indent The two-filter strategy of our current \textit{HST} survey will measure the location of the RC on the color-magnitude diagrams (CMD) for the fields with relatively low extinction to produce a new extinction map with an effective resolution of ${\leq}\,1$ arcsecond within the GBTDS footprint (excluding gaps between \textit{HST} pointings). One potential drawback of this extinction map is that of the \textit{HST} $V$-band observations at the lowest latitudes: the overall increased extinction at these latitudes [$b \lesssim -1.0$] could make it difficult to reliably detect the location of the RC due to severe differential reddening. A forthcoming study will combine these two-filter \textit{HST} observations with five additional \textit{HST}-WFC3IR near-infrared passbands that overlap with approximately 17 \textit{HST} pointings from our current survey. These accompanying \textit{HST} observations are being taken as part of GO-17173 \citep{nataf:2022prop} and GO-17923 \citep{nataf:2025prop}. The full, combined extinction analysis will be presented in \cite{nataf:inprep}.

\begin{figure*}[!htb]
\includegraphics[width=0.7\linewidth]{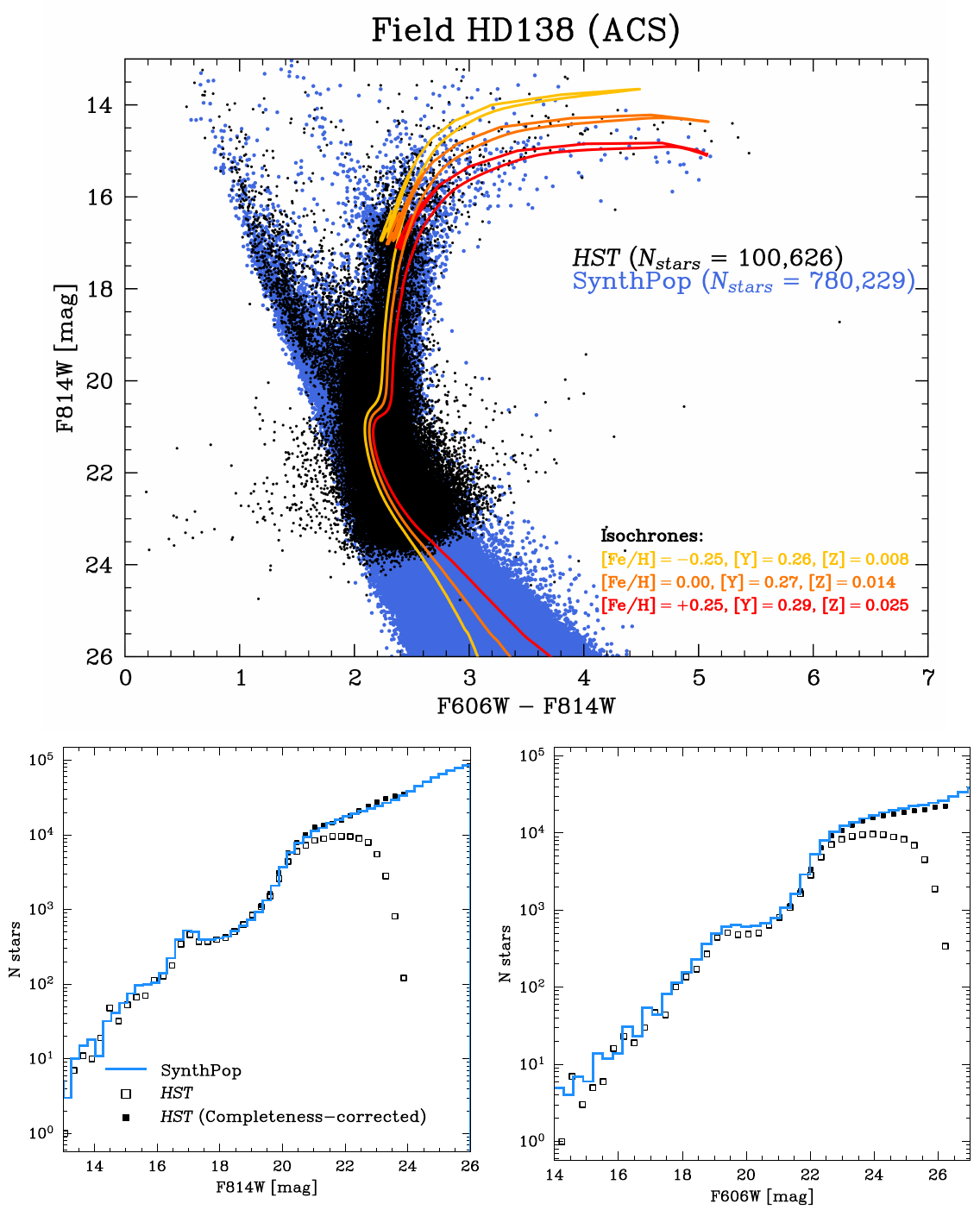}
\centering
\caption{\textit{Top panel}: Color-magnitude diagram for stars in the \textit{HST}-ACS field HD138 (black points) compared to a simulated color-magnitude diagram for the same FoV generated via \texttt{SynthPop} (blue points). We show three \texttt{MIST} stellar isochrones with varying metallicities and mass fractions (yellow, orange, red) for comparison. \textit{Bottom panels}: The observed \textit{HST} LFs (open squares) for this ACS field in the $F814W$ (\textit{Left}) and $F606W$ (\textit{Right}) passbands, compared to the \texttt{SynthPop} simulation (blue histograms). The completeness-corrected \textit{HST} counts, which come from artificial star tests (Section \ref{subsec:art-star-tests}), are shown as filled squares.}
\label{fig:CMD_LF}
\end{figure*}


\section{Field Selection and Observing Strategy} \label{sec:obs_reductions}

The field selection for this \textit{HST} survey is straightforward. Given the ROTAC-recommended GBTDS footprint shown in Figure \ref{fig:mosaic} (black outlines), we fill as much of the five contiguous \textit{Roman} fields as we can with parallel imaging from the \textit{HST} Advanced Camera for Surveys (ACS) Wide Field Camera (WFC) and Wide Field Camera 3 (WFC3) UVIS camera full-frame detectors. The prime camera is ACS/WFC and the parallel camera is WFC3/UVIS. One of several reasons we select ACS as the prime instrument is as follows; throughout the galactic bulge \textit{HST} observability season, the orientation restriction naturally results in a steady rotation of the focal plane. We can minimize the impact of \textit{HST} field overlaps by maintaining that the central pointing will always be referenced to the center of the larger FoV ACS detector. The overall \textit{HST} mosaic/tile pattern is optimized to cover the largest possible region of the GBTDS fields within the allotted number of orbits (see Table \ref{tab:survey-params} and Figure \ref{fig:mosaic}). Due to the recent \textit{HST} transition to reduced gyro mode as well as schedulability considerations, we impose no orientation, roll angle, or other observability constraints. This will inevitably result in minor overlap for a subset of adjacent WFC3 pointings, but we expect the effect to be at the few percent level considering a majority of the survey was conducted relatively quickly (between mid-February and early-June 2025).\\
\indent The survey includes a total of 177 visits with \textit{HST}, and each visit utilizes parallel-imaging mode. Therefore a total of 354 unique fields (177 ACS $+$ 177 WFC3) will be imaged by the end of the survey. Each visit consists of one orbit of \textit{HST}, with four ACS exposures and four WFC3 exposures obtained for a total of eight exposures during each orbit. The first two exposures in each camera (first four total exposures) are obtained with the $F814W$ filter. Halfway through the orbit the filter is changed on both cameras, and the remaining four total exposures are then obtained with the $F606W$ filter before the end of the orbit. The exposure times are different for the two cameras because the larger format ACS camera has a significantly longer overhead and buffer readout time compared to the WFC3 camera. The total exposure times for the ACS-$F814W$ filter is 390 sec, ACS-$F606W$ filter is 680 sec, WFC3-$F814W$ filter is 440 sec, and WFC3-$F606W$ filter is 705 sec. We note that the ACS/WFC detector is more sensitive in the red, and the WFC3/UVIS detector is more sensitive in the blue. Further details on our specific filter selections can be found in Section \ref{subsec:filters}.\\
\indent Generally, we expect this dataset to serve as a long-term Legacy \textit{HST} resource. The entire set of exposures that make up the full mosaic are being reduced with PSF fitting algorithms (Section \ref{subsec:pipeline}) to produce High-Level Science Product (HLSP) photometric and astrometric object catalogs to be hosted on the Mikulski Archive for Space Telescopes (MAST). The full potential of this \textit{HST} dataset will only be realized when the \textit{Roman} GBTDS is underway, and the \textit{Roman} and \textit{HST} datasets can be combined. However, as mentioned in Section \ref{sec:precursor-lens-detections}, there are many precursor studies that can be conducted now to enable a stronger understanding of the provisional fields and better forward modeling of the GBTDS as a whole. In Figure \ref{fig:hd70_composite} we show a color composite of the parallel field HD70 which was observed on May 9, 2025. The WFC3/UVIS position angle (PA) is rotated by ${\sim}45\degree$ with respect to ACS/WFC. The center of each image is separated by ${\sim}6$ arcminutes on the sky. As we describe in Section \ref{subsec:dithering}, partial chip gaps can be seen in both full-frame images, due to the large dither size in both X and Y detector directions. Finally, in Sections \ref{sec:first-look} and \ref{subsec:synthpop} we describe our early reduction and analysis of several parallel fields that have been imaged during the 2025 portion of the survey.

\subsection{HST Filter Selection} \label{subsec:filters}
The ACS and WFC3 detectors on \textit{HST} have filters that cover a similar region of the electromagnetic spectrum. ACS can observe wavelengths between 0.35 $\mu$m $-$ 1.1 $\mu$m, and WFC3 can observe between 0.20 $\mu$m $-$ 1.0 $\mu$m. As mentioned earlier, WFC3 has a near-infrared channel that covers wavelengths between 0.80 $\mu$m $-$ 1.7 $\mu$m, however the FoV of WFC3/IR is significantly smaller than WFC3/UVIS. Since a top priority of this survey is to cover as much sky area as possible, we have opted to exclusively use the WFC3/UVIS channel. We utilize the optical channel of the ACS/WFC instrument. Lastly, this survey is purely imaging, there are no spectroscopic observations being taken. \\
\indent We choose to utilize the $F814W$ (wide $I$-band, $\lambda_c = 0.835 \mu$m, $\Delta_{\lambda} = 0.256 \mu$m) filter as it is a well-understood passband in both cameras and has a well-characterized point spread function (PSF) \citep{anderson:2006a}. This passband is also similar to the \textit{Roman} F087 (wide $I$-band, $\lambda_c = 0.869 \mu$m, $\Delta_{\lambda} = 0.217 \mu$m) filter. Our second filter selection is $F606W$ (wide $V$-band, $\lambda_c = 0.596 \mu$m, $\Delta_{\lambda} = 0.234 \mu$m) as it is similar to the \textit{Roman} F062 (wide $V$-band, $\lambda_c = 0.620 \mu$m, $\Delta_{\lambda} = 0.280 \mu$m) filter. As we show in Figure \ref{fig:filters}, these ``bluer" \textit{HST} filters complement the primary \textit{Roman} filters that will be used during the GBTDS. Lastly, as described in Section \ref{sec:precursor-lens-detections}, an independent measurement of the source and lens color is of particular importance for determining the source radius crossing time and eventually the angular Einstein radius, $\theta_E$. Our selection of bluer ($F606W$) and redder ($F814W$) filters improves the color baseline and improves our ability to measure color-dependent centroid shifts if they are present for both historical microlensing events (Section \ref{sec:precursor-lens-detections}) and future \textit{Roman}-detected microlensing events.

\begin{deluxetable}{lcccc}
\deluxetablecaption{Early fields and their properties\label{tab:early-fields}}
\tablecolumns{6}
\setlength{\tabcolsep}{8pt}
\tablewidth{0.9\columnwidth}
\tablehead{
\colhead{Field} &
\colhead{$\ell$} & \colhead{$b$} & \colhead{$A_I$} & \colhead{$E(V-I)$}\\
\colhead{} & \colhead{[deg]} & 
\colhead{[deg]} & \colhead{[mag]} & \colhead{[mag]}
}
\startdata
HD16 (ACS) & 0.139 & -1.789 & 2.060 & 1.734\\
HD16 (WFC3) & 0.165 & -1.693 & 2.079 & 1.717\\
HD70 (ACS) & 0.829 & -1.684 & 1.554 & 1.225\\
HD70 (WFC3) & 0.863 & -1.609 & 1.724 & 1.425\\
HD98 (ACS)$^*$ & 0.972 & -1.384 & 2.786 & 2.425\\
HD98 (WFC3)$^*$ & 1.001 & -1.291 & 2.811 & 2.581\\
HD138 (ACS) & -0.496 & -1.794 & 2.464 & 2.056\\
HD138 (WFC3) & -0.459 & -1.702 & 2.447 & 2.103
\enddata
\tablenotetext{}{\footnotesize{Extinction and reddening are estimated using \cite{nataf:2013a}.\\
$^*$HD98 is at the lowest-latitude edge of the \cite{nataf:2013a} extinction map.}}
\end{deluxetable}

\subsection{HST Dither Strategy} \label{subsec:dithering}
As mentioned previously, one of the primary goals of this \textit{HST} survey is to maximize sky area coverage of the \textit{Roman} GBTDS footprint. Therefore we opted for a large-dither strategy in both the ACS and WFC3 observations. Since ACS is the primary camera for this program, we define the origin of each fields pointing to be the center of the \texttt{WFCENTER} aperture. For all 354 fields in the survey, there are a total of four exposures and three dithers between them. We do not expect to achieve extremely high-precision astrometry of the stars via many sub-pixel dithered exposures to sample the PSF extremely well. To do this would take approximately an order of magnitude more observing time (or $>1800$ \textit{HST} orbits) and would be prohibitively expensive.\\
\indent For all visits in the program we define the first exposure target position (POS TARG) as $(X,Y) = [0.0,0.0]$. The subsequent dithers between exposures one and two, two and three, and three and four are $(X,Y) = [-30.0,40.0]$ arcsec, $(X,Y) = [0.159,-0.050]$ arcsec, and $(X,Y) = [29.841,-39.950]$ arcsec, respectively. We note that stars closer to the edges of the full-frame images will be observed in only one exposure for each passband. This means, while they can have an astrometric error estimated through the multi-passband observation, they may not necessarily have a photometric error reported. While \texttt{hst1pass} automatically assigns an error of `9.990' for these stars with only a single photometric measurement, we have chosen to mask out (i.e. set empty) these error entries in the final output catalog (see Table \ref{table:sample-catalog}). We provide more detail on these stars in Section \ref{subsec:pipeline}.\\
\indent Lastly, very early in the survey we noticed a relatively high failure rate during many of the \textit{HST} visits. This impacted 26 visits, which is nearly 50\% of all visits within the first month of the survey. A vast majority of these failures (80\%) were due to a loss of guide star lock within the fine guiding system, the remaining 20\% of failures were due to guide star acquisition failure at the beginning of an orbit. While we did not investigate these failures in extreme detail, we generally attribute them to lowered capability in reduced gyro mode as we described in Section \ref{sec:obs_reductions}. Ultimately, \textit{HST} engineers modified the guide star acquisition and fine guiding scheme to significantly reduce the failure rate of subsequent visits to nearly 0\%. \textit{Hubble} Observation Problem Reports (HOPRs) were filed for each failed visit or partially failed visit, and re-visits were successfully conducted for all of these prior failures. Particularly for the partially failed visits, it is worth noting that the filter and dither combination are naturally altered since the re-visits were obtained at slightly different orientations compared to the original observations. Our reduction scheme accounts for this at the image registration stage. In total ${\sim}$6\% of the surveyed fields have these dithered observations taken at slightly different roll angles.


\section{Early Results} \label{sec:first-look}

\subsection{Data Reduction Pipeline} \label{subsec:pipeline}
\begin{figure*}[!htb]
\includegraphics[width=0.9\linewidth]{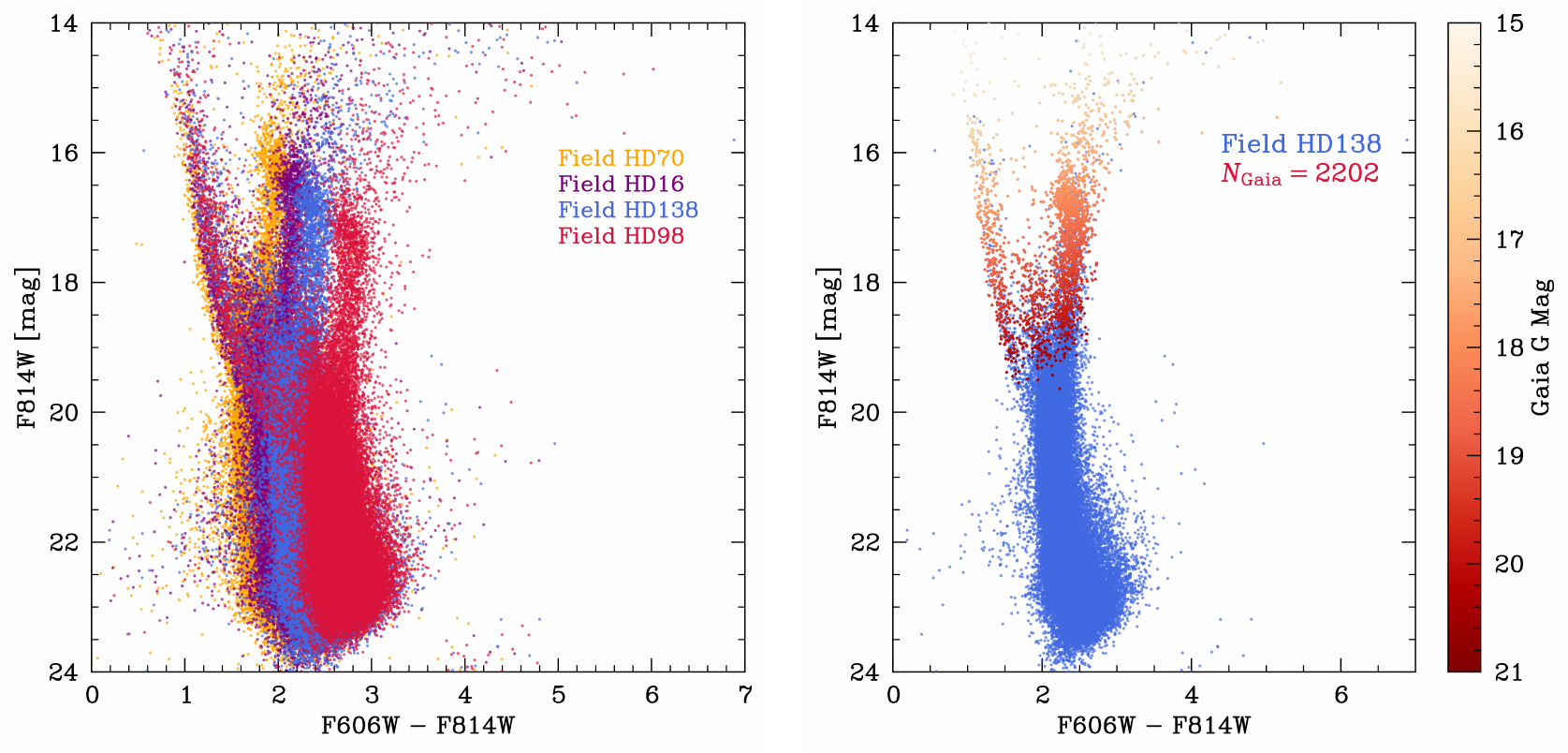}
\centering
\caption{\textit{Left panel:} Color-magnitude diagrams from four HST/WFC3 fields observed early in the survey. The field HD70 (orange points), was observed in a lower-extinction area of the footprint (see Figure \ref{fig:mosaic}). Field HD98 (red points) was observed at a higher-extinction sight line, closer to the galactic mid-plane. The variation (gradient) in extinction and reddening is apparent between the fields. \textit{Right panel:} Color-magnitude diagram for the field HD138 (blue points), with 2202 matched \textit{Gaia} DR3 stars colored by \textit{Gaia} $G-$band magnitude.}
\label{fig:joint_cmds}
\end{figure*}

We have developed a custom data reduction pipeline that is largely based on the \texttt{hst1pass} algorithm of \cite{anderson:2006a} and \cite{anderson:2022a}. We only briefly describe the reduction procedure here, and refer the reader to \cite{anderson:2022a} for a full accounting. The routine reads three things as input: (1) an \textit{HST} image (either ACS or WFC3, and either \texttt{flt} or \texttt{flc} file extension), (2) a PSF, and (3) a distortion solution (for either ACS or WFC3). Then, a pixel-by-pixel search through the image is performed, and a list of found stars with their photometric and astrometric measurements is returned. Our observing strategy described in Section \ref{sec:obs_reductions} means that a starlist is returned for both images in each passband, retaining the inherently large dither between the images. The two star lists (in each passband) are then collated into a single star list, where common stars that appear in both star lists (e.g. overlapping portion of the images) have their mean photometry and errors reported. In order to calculate a typical RMS, more than one measurement is required, so clearly stars found in only one image do not pass this simple criteria. However, the photometry measurements for these single-image stars are still important, and thus are retained in the collated star catalogs. Finally, the instrumental photometry is transformed to the VEGAmag system using \texttt{PySynphot} \citep{astropy:2013, astropy:2018}. \texttt{PySynphot} includes a proper accounting of the WFC3/UVIS chip-based variations in the zero-point values (UVIS-1 vs. UVIS-2 chip). Magnitudes we report throughout this work are thus in the VEGAmag system, including the data shown in Table \ref{table:sample-catalog} and Figures \ref{fig:CMD_LF} and \ref{fig:joint_cmds}. \\
\indent We establish the astrometric reference frame using well-measured \textit{Gaia} DR3 sources. When determining if a \textit{Gaia} star is ``well-measured", we refer to both its renormalized unit weight error (RUWE) and astrometric excess noise significance (AENS). Generally, these two metrics are used as a statistical criterion to determine the astrometric data quality in Gaia \citep{lindegren:2012a}. Based on the \textit{Gaia} DR3 documentation \citep{vallenari:2023a} and past literature \citep{stassun:2021a, lam:2022a, terry:2023a}, we remove \textit{Gaia} stars with RUWE $> 1.3$ and AENS $> 2$ mas. The alignment to the remaining well-measured \textit{Gaia} sources follows a typical 2D polynomial transformation:

\begin{equation} \label{eq:x_affine}
    x^{\prime} = a_0 + a_1x + a_2y + a_3x^2 + a_4xy + a_5y^2
\end{equation}

\begin{equation} \label{eq:y_affine}
    y^{\prime} = b_0 + b_1x + b_2y + b_3x^2 + b_4xy + b_5y^2
\end{equation}

These are affine transformations which are used to model translation, rotation, scaling, and shearing introduced by the different cameras. We perform a second pass of this fitting, where the \textit{HST} stars are aligned to themselves using a 2D polynomial going up to second order. In the second pass, the \textit{HST} catalogs are aligned to the output reference frame that was derived in the first pass, which further refines the overall reference frame. The results from this fitting scheme can vary from field to field, depending on factors like the severity of extinction and the number of well-measured \textit{Gaia} sources. The minimum number of well-measured \textit{Gaia} sources in any of our current ACS or WFC3 fields is 150, the average number of well-measured \textit{Gaia} stars per \textit{HST} field is ${\sim}$400. The typical astrometric residuals from our fitting and transformation scheme are of order ${\sim}$5 mas for ACS and ${\sim}$4 mas for WFC3. This is slightly larger than the typical astrometric uncertainties for the brighter stars measured by \texttt{hst1pass} (3 mas at $F814W\sim 19.1$, see Table \ref{tab:survey-params}).\\
\indent Table \ref{tab:early-fields} reports four fields observed early in the survey, along with interstellar extinction and reddening estimates from \cite{nataf:2013a}. We do not perform an extensive extinction/reddening analysis using the \textit{HST} data itself in the current paper. The upcoming works of \cite{nataf:inprep} and \cite{terry:inprep} will provide new details on extinction and reddening maps across the full \textit{HST} survey footprint. Table \ref{table:sample-catalog} shows an abbreviated version of the final HD138 catalog. As described earlier, this catalog delivers the distortion-corrected \textit{HST} positions of stars, their magnitudes in various \textit{HST} and \textit{Gaia} (where available) passbands, and additional information from \textit{Gaia} (parallaxes, proper motions, etc.) as well as the WFC3/IR measured values from \cite{nataf:inprep} where available. The full catalog with all early fields (Table \ref{tab:early-fields}) is available in machine-readable format online. Ultimately, a collated catalog of all measured stars from the full survey will be published in a forthcoming paper \citep{terry:inprep}.

\subsection{Artificial Star Tests} \label{subsec:art-star-tests}
Artificial star tests are useful for characterizing the photometric completeness in a stellar point source catalog. They are also helpful in identifying any significant errors in the reduction scheme, like the recovery of bright stars in the non-linear regime or finding fainter stars near the PSF wings of bright neighbors.\\
\indent We conduct injection/recovery tests on the HD138 ACS/WFC images using an artificial star mode within the Kitchen Sink 2 (\texttt{KS2}) code of \cite{anderson:2008a} to accept an input list of artificial star positions, magnitudes, and colors. The colors and magnitudes of fake stars were estimated by calculating the loci of each point along
the MS of the real CMD including Gaussian noise around each source. Artificial stars were added to each image one by one and in a “tile by tile” pattern, adding and measuring synthetic stars in each tile of size ${\sim}$120 × 120 pixels. This method is useful for avoiding major effects from crowding, which can be substantial in galactic bulge field images like HD138. The output art-star positions and fluxes were then compared with the input art-star files. A star was considered found if it passed the following criteria:

\begin{equation}\label{eq:art-star_positions}
    \sqrt{(X_{\textrm{out}} - X_{\textrm{in}})^2 + (Y_{\textrm{out}} - Y_{\textrm{in}})^2} \leq 0.50\textrm{pix},
\end{equation}

\begin{equation}\label{eq:art-star_mags}
    \left| \textrm{mag}_{\textrm{in}} - \textrm{mag}_{\textrm{out}} \right| \leq 0.50\textrm{mag},
\end{equation}

\noindent where `in' and `out' denote input star and output stars respectively. The bottom panels of Figure \ref{fig:CMD_LF} show the LFs for the stars in the HD138 (ACS) field. The observed (uncorrected) LFs are shown as black open squares in the $F814W$ (left) and $F606W$ (right) passbands, and the completeness-corrected LFs are shown as black filled squares. The completeness-corrected counts are in reasonable agreement with a \texttt{SynthPop} synthetic population of stars along the same sight line (blue histograms). We describe our comparison to the \texttt{SynthPop} catalog in more detail in the next subsection. The raw empirical star counts in this HD138 dataset become measurably incomplete around $F814W\,{\sim}\,19.5$ mag ($F606W\,{\sim}\,22$ mag) and reach a completeness of 50\% at $F814W\,{\sim}\,22.3$ mag ($F606W\,{\sim}\,23.8$ mag). At present we have performed artificial star tests on only the HD138 field, as this paper is largely an overview and early results work. Although we do not expect significant variations, in subsequent work \citep{terry:inprep} we will measure the completeness via injection/recovery tests across several additional ACS/WFC3 fields in the survey, spanning a wide range of on-sky locations.

\subsection{Comparing the Empirical Catalog to Population Synthesis Models} \label{subsec:synthpop}
With an early reduction and generation of a star catalog using our pipeline, we compare the stellar density (e.g. the LF) and CMD between our empirical star catalog and a synthetic star catalog generated by the \texttt{SynthPop} Galactic model (\citetalias{kluter:2025a}, \citeyear{kluter:2025a}). The \texttt{SynthPop} framework uses a Galactic density model, which includes the bulge, thin and thick disks, nuclear stellar disk, and stellar halo. The primary components are based on microlensing survey results \citep{nataf:2013a, udalski:2015a, mroz:2019a}, with the bulge density model from \citet{cao:2013a} and disks from \citet{koshimoto:2021a}. The bulge and disk kinematic models are pulled from \cite{koshimoto:2021a} and halo density and kinematics from \cite{robin:2003a}. The nuclear stellar disk density and kinematics are drawn from \citet{sormani:2022a} and contribute minimally ($<1$\%) to the simulated stellar contents in the five contiguous GBTDS fields. For all populations, the model adopts the \citet{kroupa:2001a} initial mass function and MIST \citep[MESA Isochrones and Stellar Tracks]{choi:2016a, dotter:2016a} isochrones. For extinction, the model uses the 2-d map of \citet{Surot:2020a}, distributed along the line of sight according to the double exponential disk scaling method of \citet{sharma:2011a}. The extinction law comes from \citet{Cardelli:1989a}, with the optical side set by \citet{ODonnell:1994a} and infrared side adjusted to match \citet{Surot:2020a}. A forthcoming paper by \cite{huston:inprep} will present additional details of the model that we've implemented in this work. \\
\indent The top panel of Figure \ref{fig:CMD_LF} shows the CMD for the \textit{HST}-measured stars in the HD138 ACS field (black points), along with the synthetic stars that are drawn from the \texttt{SynthPop} Galactic model toward the same sight line (blue points). For comparison, we plot three \texttt{MIST} isochrones in yellow, orange, and red solid lines. When generating these isochrones, we keep the stellar age constant at 10 Gyr and generate sub-solar ([Fe/H] $=$ $-0.25$), [Y] $=$ 0.26, [Z] $=$ 0.008), solar-like ([Fe/H] $=$ $0.0$), [Y] $=$ 0.27, [Z] $=$ 0.014), and super-solar ([Fe/H] $=$ $0.25$), [Y] $=$ 0.29, [Z] $=$ 0.025) compositions. [Fe/H] is a measure of the iron abundance, [Y] represents the helium fraction, and [Z] represents the metallicity. As described in Section \ref{subsec:art-star-tests}, the \textit{HST} star counts are near fully complete until at or just below the main sequence turnoff (MSTO) in this field.

\subsection{Cross-Referencing with \textit{Gaia} DR3}
As discussed in Section \ref{subsec:pipeline}, we cross-match our \textit{HST} catalogs to well-measured \textit{Gaia} DR3 sources in order to establish the astrometric reference frame. Although we impose a somewhat stringent cutting criteria for the \textit{Gaia} stars selected for the coordinate transformation, we do retain all cross-matched sources and include them in the final star catalogs for each field.\\
\indent Clearly, the limiting magnitude of \textit{Gaia}, particularly in the very crowded bulge region, prohibits most of the \textit{HST} stars from having a \textit{Gaia}-measured counterpart. Further, the level of extinction along different lines of sight also effects the total number of \textit{Gaia} sources we are able to match. We find that, on average we can successfully match ${\sim 2,300}$ \textit{Gaia} stars per ACS/WFC field and ${\sim 1,300}$ \textit{Gaia} stars per WFC3/UVIS field. We find that the number of matched \textit{Gaia} stars typically accounts for ${\sim}75$\% of the similar brightness range \textit{HST} stars we detect with \texttt{hst1pass} in our fields, down to the completeness of \textit{Gaia} ($G_{\textrm{mag}} \sim 20$). Since our \textit{HST} catalog goes much deeper than \textit{Gaia}, there is a chance for spurious matches. That is, for a single \textit{Gaia}-reported source, there may be multiple \textit{HST} sources in the vicinity. As a check for spurious matches, we examine the color-color relation between \textit{Gaia} and \textit{HST}. Any large deviation from a smoothly continuous relation would imply a spurious match (i.e. \textit{Gaia} star of a given color is matched to incorrect star in \textit{HST}). The much brighter, relatively speaking, limiting magnitude in \textit{Gaia} actually helps us avoid any major issues with spurious matching. This is because nearly all of the \textit{Gaia} sources in these fields are bright giants or sub-giants. The crowding due to giants/sub-giants is not as severe as the crowding due to stars below the MSTO in these galactic bulge fields. We find on average, a very small number ($< 0.1$\%) of spurious matches where the suspect star deviates more than 5$\sigma$ from the loci of the color-color relation. Integrated across all of the proposed \textit{Gaia} cross-matches in our early fields catalog, this is approximately 15 candidate spurious matches. For these stars in question we release the \textit{Gaia} cross-identification as a precaution. We estimate, conservatively, that by the time the full \textit{HST} catalog is generated at the end of the survey, the residual rate of spurious \textit{Gaia} DR3 cross-matches will be $\sim 0.01$\%, or about 80 out of 800,000 total cross-matched \textit{Gaia} sources. In some of the highest extinction fields, we are only able to successfully match $500 - 900$ \textit{Gaia} stars, nearly all of which are foreground disk stars and giant stars in the bulge. The right panel of Figure \ref{fig:joint_cmds} shows a moderately-extincted ($A_I = 2.79$) CMD (red points), along with all matched \textit{Gaia} DR3 sources (colored points). The color bar denotes \textit{Gaia} $G-$band magnitudes. As expected, nearly all of the matched \textit{Gaia} stars are foreground disk stars and bulge stars above the MSTO in this field.

\begin{deluxetable*}{cccccccclcc}
\deluxetablecaption{Subset of the Early Release catalog\label{table:sample-catalog}}
\tablecolumns{11}
\setlength{\tabcolsep}{5pt}
\tablewidth{\columnwidth}
\tablehead{
\colhead{$X$} & \colhead{${X_{\textrm{err}}}$} &
\colhead{$Y$} & \colhead{${Y_{\textrm{err}}}$} & \colhead{$F606W$} & \colhead{${F606W_{\textrm{err}}}$} & \colhead{$F814W$} & \colhead{${F814W_{\textrm{err}}}$} 
& \colhead{Field} & \colhead{RA} & \colhead{DEC}
}
\startdata
5060.315 & 0.053 & 3120.144 & 0.033 & 22.415 & 0.013 & 20.883 & 0.008 & HD70$_{\textrm{ACS}}$ & 268.544 & -29.110 \\
3154.281 & 0.028 & 5444.921 & 0.024 & 22.767 & 0.016 & 20.883 & 0.011 & HD16$_{\textrm{ACS}}$ & 268.269 & -29.731 \\
2528.627 & 0.032 & 4132.876 & -- & 22.962 & -- & 20.883 & -- & HD138$_{\textrm{WFC3}}$ & 267.839 & -30.212 \\
2567.233 & 0.027 & 4434.218 & 0.012 & 23.251 & -- & 20.883 & -- & HD70$_{\textrm{ACS}}$ & 268.576 & -29.095 \\
4794.341 & 0.052 & 5459.233 & 0.037 & 22.794 & 0.021 & 20.883 & 0.050 & HD16$_{\textrm{WFC3}}$ & 268.168 & -29.658 \\
2798.477 & 0.049 & 6363.276 & 0.023 & 22.467 & 0.018 & 20.883 & 0.020 & HD70$_{\textrm{ACS}}$ & 268.573 & -29.073 \\
7625.562 & 0.026 & 5748.217 & 0.051 & 22.775 & -- & 20.883 & -- & HD16$_{\textrm{ACS}}$ & 268.212 & -29.728 \\
3145.680 & 0.048 & 6013.260 & 0.019 & 22.794 & 0.010 & 20.883 & 0.034 & HD16$_{\textrm{ACS}}$ & 268.269 & -29.725 \\
6242.964 & 0.038 & 3818.430 & 0.033 & 22.829 & 0.012 & 20.883 & 0.009 & HD16$_{\textrm{WFC3}}$ & 268.149 & -29.676 \\
5628.925 & 0.013 & 7727.328 & 0.035 & 22.981 & -- & 20.883 & -- & HD16$_{\textrm{WFC3}}$ & 268.157 & -29.633 \\
2032.043 & 0.080 & 6709.665 & 0.015 & 22.530 & -- & 20.883 & -- & HD70$_{\textrm{ACS}}$ & 268.582 & -29.070 \\
6792.515 & 0.063 & 4452.601 & 0.027 & 22.561 & 0.014 & 20.883 & 0.035 & HD70$_{\textrm{ACS}}$ & 268.522 & -29.095 \\
4251.116 & 0.044 & 5340.480 & 0.015 & 23.033 & 0.035 & 20.883 & 0.010 & HD98$_{\textrm{ACS}}$ & 268.340 & -28.809 \\
4946.974 & 0.043 & 5149.807 & 0.015 & 22.737 & 0.023 & 20.883 & 0.011 & HD16$_{\textrm{WFC3}}$ & 268.166 & -29.661 \\
3394.159 & 0.019 & 6752.509 & 0.051 & 23.044 & -- & 20.883 & 0.054 & HD98$_{\textrm{ACS}}$ & 268.351 & -28.794 \\
7129.029 & 0.012 & 5745.118 & 0.013 & 23.373 & -- & 20.883 & -- & HD16$_{\textrm{WFC3}}$ & 268.138 & -29.655
\enddata
\tablenotetext{}{\footnotesize{The full catalog of 767,415 sources is available in machine-readable format online, and includes additional columns from \textit{Gaia} where available.}}
\end{deluxetable*}

\subsection{Cross-Referencing with the Euclid Galactic Bulge Survey} \label{subsec:euclid-roman}
On March 23 2025, \textit{Euclid} obtained high-resolution images covering 4.8 deg$^2$ of the inner bulge south of the GC (including the GBTDS footprint) with the \textit{VIS} Camera. Compared to the regular \textit{Euclid} observing sequence, the exposure time was slightly shorter (400 sec instead of 550 sec), with 16 dithers at each of the nine total pointings to ensure a well-sampled PSF. This program is known as the \textit{Euclid} Galactic Bulge Survey\footnote{\url{https://www.cosmos.esa.int/web/euclid/egbs}}. Additional calibration fields were observed under the same thermal environment before and after the \textit{Euclid} Galactic Bulge Survey, which ensures a proper PSF modeling. The data is currently undergoing processing by the \textit{Euclid} Exoplanet Science Working Group, with a scheduled public release in June 2026.\\
\indent The inclusion of this \textit{Euclid} Galactic Bulge Survey precursor dataset will synergize well with our \textit{HST} survey, especially for the historical microlensing events. The added benefit of an additional high-resolution dataset is also apparent when considering the future microlensing events that \textit{Roman} will detect during the GBTDS, which will be particularly helpful for lens mass and distance measurements. A joint analysis of the well-dithered \textit{Euclid} VIS images with the multi-passband \textit{HST} images can raise the number of \textit{Roman}-detected microlensing events that have successful lens detections and relative proper motion measurements. Since the lens mass measurement precision is sensitive to the cube of the \textit{Euclid/HST} and \textit{Roman} observation baseline for partially blended lens-source pairs (i.e. $\Delta x^{-3} \propto \Delta t^{-3}$), we expect a ${\sim}3\times$ improvement in the accuracy of \textit{Roman} mass measurements for events with lens-source relative proper motions $\mu_{\textrm{rel}} > 9$ mas yr$^{-1}$ (E. Kerins private communication).\\
\indent Lastly, \cite{Thygesen:inprep} are preparing a multi-band ground-based photometry catalog covering the \textit{HST/Roman/Euclid} footprints. They have reprocessed $J,H,K$ images from the VISTA Variables in the Via Lactea (VVV) survey \citep{minniti:2010a} for all overlapping fields. These VVV catalogs are combined with CFHT-derived catalogs obtained in $U,G,R,I$, as well as $V,I$ photometry from the OGLE survey. This
will ultimately provide multi-band coverage for the brightest stars in the fields, and it is expected that this catalog will be released in June 2026, along with the \textit{Euclid} public data release.


\section{Discussion and Conclusion} \label{sec:conclusion}

We have presented a wide-field \textit{HST} survey of the central Galactic bulge region. The survey area was strategically selected to coincide with five contiguous fields that will be observed by the upcoming \textit{Roman} GBTDS. One of the primary goals of this \textit{HST} survey is to prepare for \textit{Roman} by measuring, with high spatial resolution, sources and lenses several years before they participate in gravitational microlensing events that will be detected during the \textit{Roman} GBTDS. Since we do not know the exact location of future \textit{Roman}-detected events, our \textit{HST} survey strategy is optimized for maximal FoV coverage of the \textit{Roman} footprint. To aid in this effort we utilize parallel-imaging mode to conduct simultaneous observations with the full-frame ACS/WFC and WFC3/UVIS cameras. \\
\indent Further, we have discussed several additional science cases that this survey addresses, such as late-time followup analyses of historical microlensing events that have been detected from the ground, and high-resolution extinction and reddening maps across these patchy and highly-variable Galactic bulge sight lines. Additionally, we have outlined prospects to utilize the complete \textit{HST} catalog from this survey as a high-fidelity input dataset to seed the eventual \textit{Roman} GBTDS catalog. \\
\indent We developed a custom data reduction pipeline that performs PSF photometry and astrometry on all images and passbands, with photometry calibrated to the Vega magnitude system and astrometry transformed to an absolute reference frame using \textit{Gaia} DR3. We present early results from our analysis of four fields that show varying extinction, reddening, and stellar surface density. We have performed a specialized analysis of the HD138 field, where we compare the empirical star counts, CMD, and LF to a population synthesis sample drawn from \texttt{SynthPop} as well as stellar isochrone models from \texttt{MIST}. Artificial star tests in this field give us completeness-corrected stellar surface densities that are consistent with the \texttt{SynthPop} sample. The calibrated catalogs for these four early fields (Table 2) are available in machine-readable format, and the full 354-field catalog will be released in \cite{terry:inprep}. \\
\indent Ultimately, we aim for this \textit{HST} survey to be largely supportive of, and enhance in many ways, the science output of the \textit{Roman} GBTDS. The final \textit{HST} catalog of ${\sim}25$ million stars will benefit future bulge population studies, dynamics, exoplanet systems, extinction, metallicities and much more as a legacy dataset for the community as we advance toward the \textit{Roman} era and beyond.

\begin{acknowledgements}
Support for program GO-17776 was provided by NASA through a grant from STScI, which is operated by AURA, Inc., under NASA contract NAS 5-26555. SKT, JA, CAB, DPB, AB, BSG, MJH, JRL, DMN, MTP, and AV are supported by the Nancy Grace Roman Space Telescope Project through the National Aeronautics and Space Administration grant 80NSSC24M0022. BSG is also supported by The Ohio State University through the Thomas Jefferson Chair for Discovery and Space Exploration endowment. JPB and NR have been supported by the Australian Government through the Australian Research Council Discovery Project Grant 240101842 and the SPACE-MLENS ANR grant ANR-24-CE31-3263. MJH Acknowledges support from the Heising-Simons Foundation under grant No. 2022-3542. All of the data presented in this paper were obtained from the Mikulski Archive for Space Telescopes (MAST) at the Space Telescope Science Institute. The specific observations analyzed can be accessed via doi:\dataset[10.17909/fq12-f295]{10.17909/fq12-f295}.
\end{acknowledgements}

\textit{Software}: Astropy \citep{astropy:2013, astropy:2018, astropy:2022}, hst1pass \citep{anderson:2022a}, KS2 \citep{anderson:2008a}, Matplotlib \citep{hunter:2007a}, Numpy \citep{harris:2020array}, SynthPop (\citetalias{kluter:2025a}, \citeyear{kluter:2025a})

\bibliographystyle{aasjournal}
\bibliography{Terry_HST_Roman.bib}

@article{kozlowski:2007a,
  title={The first direct detection of a gravitational $\mu$-lens toward the Galactic bulge},
  author={Koz{\l}owski, S and Wo{\'z}niak, PR and Mao, S and Wood, A},
  journal={The Astrophysical Journal},
  volume={671},
  number={1},
  pages={420--426},
  year={2007}
}

@article{alcock:2001a,
  title={Direct detection of a microlens in the Milky Way},
  author={Alcock, C and Allsman, RA and Axelrod, TS and Bennett, DP and Drake, AJ and Freeman, KC and Geha, M and others},
  journal={Nature},
  volume={414},
  number={6864},
  pages={617--619},
  year={2001},
  publisher={Nature Publishing Group UK London}
}

@article{haywood:2016a,
  title={Hiding its age: the case for a younger bulge},
  author={Haywood, M and Di Matteo, P and Snaith, O and Calamida, A},
  journal={Astronomy \& Astrophysics},
  volume={593},
  pages={A82},
  year={2016},
  publisher={EDP Sciences}
}

@article{pietrukowicz:2015a,
  title={Deciphering the 3D structure of the old galactic bulge from the OGLE RR Lyrae stars},
  author={Pietrukowicz, P and Koz{\l}owski, S and Skowron, J and Soszy{\'n}ski, I and Udalski, A and Poleski, R and Wyrzykowski, {\L} and Szyma{\'n}ski, MK and Pietrzy{\'n}ski, G and Ulaczyk, K and others},
  journal={The Astrophysical Journal},
  volume={811},
  number={2},
  pages={113},
  year={2015},
  publisher={The American Astronomical Society}
}

@article{kunder:2012a,
  title={The bulge radial velocity assay (BRAVA). II. Complete sample and data release},
  author={Kunder, Andrea and Koch, Andreas and Michael Rich, R and De Propris, Roberto and Howard, Christian D and Stubbs, Scott A and Johnson, Christian I and Shen, Juntai and Wang, Yougang and Robin, Annie C and others},
  journal={The Astronomical Journal},
  volume={143},
  number={3},
  pages={57},
  year={2012},
  publisher={The American Astronomical Society}
}

@article{portail:2016a,
  title={Dynamical modelling of the galactic bulge and bar: the Milky Way’s pattern speed, stellar and dark matter mass distribution},
  author={Portail, Matthieu and Gerhard, Ortwin and Wegg, Christopher and Ness, Melissa},
  journal={Monthly Notices of the Royal Astronomical Society},
  pages={stw2819},
  year={2016},
  publisher={Oxford University Press}
}

@article{portail:2015a,
  title={Made-to-measure models of the Galactic box/peanut bulge: stellar and total mass in the bulge region},
  author={Portail, Matthieu and Wegg, C and Gerhard, O and Martinez-Valpuesta, I},
  journal={Monthly Notices of the Royal Astronomical Society},
  volume={448},
  number={1},
  pages={713--731},
  year={2015},
  publisher={Oxford University Press}
}

@article{kent:1992a,
  title={Galactic structure from the spacelab infrared telescope. III-A dynamical model for the Milky Way bulge},
  author={Kent, SM},
  journal={Astrophysical Journal, Part 1 (ISSN 0004-637X), vol. 387, March 1, 1992, p. 181-188.},
  volume={387},
  pages={181--188},
  year={1992}
}

@article{sellwood:1988a,
  title={A maximum disc model for the galaxy},
  author={Sellwood, JA and Sanders, RH},
  journal={Monthly Notices of the Royal Astronomical Society},
  volume={233},
  number={3},
  pages={611--620},
  year={1988},
  publisher={Oxford University Press Oxford, UK}
}

@article{bensby:2017a,
  title={Chemical evolution of the Galactic bulge as traced by microlensed dwarf and subgiant stars-VI. Age and abundance structure of the stellar populations in the central sub-kpc of the Milky Way},
  author={Bensby, T and Feltzing, Sofia and Gould, A and Yee, Jennifer C and Johnson, JA and Asplund, Martin and Mel{\'e}ndez, J and Lucatello, Sara and Howes, LM and McWilliam, Andrew and others},
  journal={Astronomy \& Astrophysics},
  volume={605},
  pages={A89},
  year={2017},
  publisher={EDP Sciences}
}

@article{hambleton:2023a,
  title={Rubin observatory LSST transients and variable stars roadmap},
  author={Hambleton, Kelly M and Bianco, Federica B and Street, Rachel and Bell, Keaton and Buckley, David and Graham, Melissa and Hernitschek, Nina and Lund, Michael B and Mason, Elena and Pepper, Joshua and others},
  journal={Publications of the Astronomical Society of the Pacific},
  volume={135},
  number={1052},
  pages={105002},
  year={2023},
  publisher={The Astronomical Society of the Pacific}
}

@article{gould:2014b,
  title={WFIRST ultra-precise astrometry I: Kuiper belt objects},
  author={Gould, Andrew},
  journal={arXiv preprint arXiv:1403.4241},
  year={2014}
}

@article{weiss:2025a,
  title={Modeling Asteroseismic Yields for the Roman Galactic Bulge Time-domain Survey},
  author={Weiss, Trevor J and Downing, Noah J and Pinsonneault, Marc H and Zinn, Joel C and Stello, Dennis and Bedding, Timothy R and Cao, Kaili and Hon, Marc and Reyes, Claudia and Gaudi, B Scott and others},
  journal={The Astrophysical Journal},
  volume={987},
  number={2},
  pages={181},
  year={2025},
  publisher={The American Astronomical Society}
}

@article{gould:2014a,
  title={WFIRST Ultra-precise astrometry II: asteroseismology},
  author={Gould, Andrew and Huber, Daniel and Penny, Matthew and Stello, Dennis},
  journal={arXiv preprint arXiv:1410.7395},
  year={2014}
}

@article{gaudi:2019a,
  title={" Auxiliary" Science with the WFIRST Microlensing Survey},
  author={Gaudi, B Scott and Akeson, Rachel and Anderson, Jay and Bachelet, Etienne and Bennett, David P and Bhattacharya, Aparna and Bozza, Valerio and Novati, Sebastiano Calchi and Henderson, Calen B and Johnson, Samson A and others},
  journal={arXiv preprint arXiv:1903.08986},
  year={2019}
}

@article{paczynski:1991a,
  title={Gravitational microlensing of the Galactic bulge stars},
  author={Paczynski, B},
  journal={Astrophysical Journal, Part 2-Letters (ISSN 0004-637X), vol. 371, April 20, 1991, p. L63-L67.},
  volume={371},
  pages={L63--L67},
  year={1991}
}

@ARTICLE{moniez:2010a,
       author = {{Moniez}, Marc},
        title = "{Microlensing as a probe of the Galactic structure: 20 years of microlensing optical depth studies}",
      journal = {General Relativity and Gravitation},
         year = 2010,
        month = sep,
       volume = {42},
       number = {9},
        pages = {2047-2074},
          doi = {10.1007/s10714-009-0925-4},
archivePrefix = {arXiv},
       eprint = {1001.2707},
 primaryClass = {astro-ph.GA},
       adsurl = {https://ui.adsabs.harvard.edu/abs/2010GReGr..42.2047M}
}

@ARTICLE{kiraga:1994a,
       author = {{Kiraga}, Marcin and {Paczynski}, Bohdan},
        title = "{Gravitational microlensing of the Galactic bulge stars}",
      journal = {\apjl},
         year = 1994,
        month = aug,
       volume = {430},
       number = {2},
        pages = {L101-L104},
          doi = {10.1086/187448},
       adsurl = {https://ui.adsabs.harvard.edu/abs/1994ApJ...430L.101K}
}

@ARTICLE{gould:2009a,
       author = {{Gould}, A. and {Udalski}, A. and {Monard}, B. and {Horne}, K. and {Dong}, Subo and {Miyake}, N. and {Sahu}, K. and {Bennett}, D.~P. and {Wyrzykowski}, {\L}. and {Soszy{\'n}ski}, I. and {Szyma{\'n}ski}, M.~K. and {Kubiak}, M. and {Pietrzy{\'n}ski}, G. and {Szewczyk}, O. and {Ulaczyk}, K. and {OGLE Collaboration} and {Allen}, W. and {Christie}, G.~W. and {DePoy}, D.~L. and {Gaudi}, B.~S. and {Han}, C. and {Lee}, C.-U. and {McCormick}, J. and {Natusch}, T. and {Park}, B.-G. and {Pogge}, R.~W. and {{\ensuremath{\mu}}FUN Collaboration} and {Allan}, A. and {Bode}, M.~F. and {Bramich}, D.~M. and {Burgdorf}, M.~J. and {Dominik}, M. and {Fraser}, S.~N. and {Kerins}, E. and {Mottram}, C. and {Snodgrass}, C. and {Steele}, I.~A. and {Street}, R. and {Tsapras}, Y. and {RoboNet Collaboration} and {Abe}, F. and {Bond}, I.~A. and {Botzler}, C.~S. and {Fukui}, A. and {Furusawa}, K. and {Hearnshaw}, J.~B. and {Itow}, Y. and {Kamiya}, K. and {Kilmartin}, P.~M. and {Korpela}, A. and {Lin}, W. and {Ling}, C.~H. and {Masuda}, K. and {Matsubara}, Y. and {Muraki}, Y. and {Nagaya}, M. and {Ohnishi}, K. and {Okumura}, T. and {Perrott}, Y.~C. and {Rattenbury}, N. and {Saito}, To. and {Sako}, T. and {Skuljan}, L. and {Sullivan}, D.~J. and {Sumi}, T. and {Sweatman}, W.~L. and {Tristram}, P.~J. and {Yock}, P.~C.~M. and {MOA Collaboration} and {Albrow}, M. and {Beaulieu}, J.~P. and {Coutures}, C. and {Calitz}, H. and {Caldwell}, J. and {Fouque}, P. and {Martin}, R. and {Williams}, A. and {PLANET Collaboration}},
        title = "{The Extreme Microlensing Event OGLE-2007-BLG-224: Terrestrial Parallax Observation of a Thick-Disk Brown Dwarf}",
      journal = {\apjl},
         year = 2009,
        month = jun,
       volume = {698},
       number = {2},
        pages = {L147-L151},
          doi = {10.1088/0004-637X/698/2/L147},
archivePrefix = {arXiv},
       eprint = {0904.0249},
 primaryClass = {astro-ph.GA},
       adsurl = {https://ui.adsabs.harvard.edu/abs/2009ApJ...698L.147G}
}

@ARTICLE{wyrzykowski:2020a,
       author = {{Wyrzykowski}, {\L}. and {Mr{\'o}z}, P. and {Rybicki}, K.~A. and {Gromadzki}, M. and {Ko{\l}aczkowski}, Z. and {Zieli{\'n}ski}, M. and {Zieli{\'n}ski}, P. and {Britavskiy}, N. and {Gomboc}, A. and {Sokolovsky}, K. and {Hodgkin}, S.~T. and {Abe}, L. and {Aldi}, G.~F. and {AlMannaei}, A. and {Altavilla}, G. and {Al Qasim}, A. and {Anupama}, G.~C. and {Awiphan}, S. and {Bachelet}, E. and {Bak{\i}{\c{s}}}, V. and {Baker}, S. and {Bartlett}, S. and {Bendjoya}, P. and {Benson}, K. and {Bikmaev}, I.~F. and {Birenbaum}, G. and {Blagorodnova}, N. and {Blanco-Cuaresma}, S. and {Boeva}, S. and {Bonanos}, A.~Z. and {Bozza}, V. and {Bramich}, D.~M. and {Bruni}, I. and {Burenin}, R.~A. and {Burgaz}, U. and {Butterley}, T. and {Caines}, H.~E. and {Caton}, D.~B. and {Calchi Novati}, S. and {Carrasco}, J.~M. and {Cassan}, A. and {{\v{C}}epas}, V. and {Cropper}, M. and {Chru{\'s}li{\'n}ska}, M. and {Clementini}, G. and {Clerici}, A. and {Conti}, D. and {Conti}, M. and {Cross}, S. and {Cusano}, F. and {Damljanovic}, G. and {Dapergolas}, A. and {D'Ago}, G. and {de Bruijne}, J.~H.~J. and {Dennefeld}, M. and {Dhillon}, V.~S. and {Dominik}, M. and {Dziedzic}, J. and {Erece}, O. and {Eselevich}, M.~V. and {Esenoglu}, H. and {Eyer}, L. and {Figuera Jaimes}, R. and {Fossey}, S.~J. and {Galeev}, A.~I. and {Grebenev}, S.~A. and {Gupta}, A.~C. and {Gutaev}, A.~G. and {Hallakoun}, N. and {Hamanowicz}, A. and {Han}, C. and {Handzlik}, B. and {Haislip}, J.~B. and {Hanlon}, L. and {Hardy}, L.~K. and {Harrison}, D.~L. and {van Heerden}, H.~J. and {Hoette}, V.~L. and {Horne}, K. and {Hudec}, R. and {Hundertmark}, M. and {Ihanec}, N. and {Irtuganov}, E.~N. and {Itoh}, R. and {Iwanek}, P. and {Jovanovic}, M.~D. and {Janulis}, R. and {Jel{\'\i}nek}, M. and {Jensen}, E. and {Kaczmarek}, Z. and {Katz}, D. and {Khamitov}, I.~M. and {Kilic}, Y. and {Klencki}, J. and {Kolb}, U. and {Kopacki}, G. and {Kouprianov}, V.~V. and {Kruszy{\'n}ska}, K. and {Kurowski}, S. and {Latev}, G. and {Lee}, C.-H. and {Leonini}, S. and {Leto}, G. and {Lewis}, F. and {Li}, Z. and {Liakos}, A. and {Littlefair}, S.~P. and {Lu}, J. and {Manser}, C.~J. and {Mao}, S. and {Maoz}, D. and {Martin-Carrillo}, A. and {Marais}, J.~P. and {Maskoli{\={u}}nas}, M. and {Maund}, J.~R. and {Meintjes}, P.~J. and {Melnikov}, S.~S. and {Ment}, K. and {Miko{\l}ajczyk}, P. and {Morrell}, M. and {Mowlavi}, N. and {Mo{\'z}dzierski}, D. and {Murphy}, D. and {Nazarov}, S. and {Netzel}, H. and {Nesci}, R. and {Ngeow}, C.-C. and {Norton}, A.~J. and {Ofek}, E.~O. and {Pak{\v{s}}tien{\.{e}}}, E. and {Palaversa}, L. and {Pandey}, A. and {Paraskeva}, E. and {Pawlak}, M. and {Penny}, M.~T. and {Penprase}, B.~E. and {Piascik}, A. and {Prieto}, J.~L. and {Qvam}, J.~K.~T. and {Ranc}, C. and {Rebassa-Mansergas}, A. and {Reichart}, D.~E. and {Reig}, P. and {Rhodes}, L. and {Rivet}, J.-P. and {Rixon}, G. and {Roberts}, D. and {Rosi}, P. and {Russell}, D.~M. and {Zanmar Sanchez}, R. and {Scarpetta}, G. and {Seabroke}, G. and {Shappee}, B.~J. and {Schmidt}, R. and {Shvartzvald}, Y. and {Sitek}, M. and {Skowron}, J. and {{\'S}niegowska}, M. and {Snodgrass}, C. and {Soares}, P.~S. and {van Soelen}, B. and {Spetsieri}, Z.~T. and {Stankevi{\v{c}}i{\={u}}t{\.{e}}}, A. and {Steele}, I.~A. and {Street}, R.~A. and {Strobl}, J. and {Strubble}, E. and {Szegedi}, H. and {Tinjaca Ramirez}, L.~M. and {Tomasella}, L. and {Tsapras}, Y. and {Vernet}, D. and {Villanueva}, S. and {Vince}, O. and {Wambsganss}, J. and {van der Westhuizen}, I.~P. and {Wiersema}, K. and {Wium}, D. and {Wilson}, R.~W. and {Yoldas}, A. and {Zhuchkov}, R. Ya. and {Zhukov}, D.~G. and {Zdanavi{\v{c}}ius}, J. and {Zo{\l}a}, S. and {Zubareva}, A.},
        title = "{Full orbital solution for the binary system in the northern Galactic disc microlensing event Gaia16aye}",
      journal = {\aap},
         year = 2020,
        month = jan,
       volume = {633},
          eid = {A98},
        pages = {A98},
          doi = {10.1051/0004-6361/201935097},
archivePrefix = {arXiv},
       eprint = {1901.07281},
 primaryClass = {astro-ph.SR},
       adsurl = {https://ui.adsabs.harvard.edu/abs/2020A&A...633A..98W}
}

@article{lastovka:2025a,
  title={Predictions of the Nancy Grace Roman Space Telescope Galactic Exoplanet Survey. III. Detectability of Giant Exomoons of Wide-separation Giant Planets},
  author={Lastovka, Matthew and Gaudi, B Scott and Johnson, Samson A and Penny, Matthew T and Kerins, Eamonn and Rattenbury, Nicholas J},
  journal={The Astronomical Journal},
  volume={170},
  number={5},
  pages={258},
  year={2025},
  publisher={The American Astronomical Society}
}

@article{saggese:2025a,
  title={Predictions of the Nancy Grace Roman Space Telescope Galactic Exoplanet Survey. V. Detection Rates of Multiplanetary Systems in High Magnification Microlensing Events},
  author={Saggese, Vito and Bachelet, {\'E}tienne and Novati, Sebastiano Calchi and Bozza, Valerio and Covone, Giovanni and Zohrabi, Farzaneh and Albrow, Michael D and Anderson, Jay and Beichman, Charles and Bennett, David P and others},
  journal={arXiv preprint arXiv:2512.05182},
  year={2025}
}

@article{sanderson:2019a,
  title={Astrometry with the wide-field infrared space telescope},
  author={WFIRST Astrometry Working Group, The and Sanderson, Robyn E and Bellini, Andrea and Casertano, Stefano and Lu, Jessica R and Melchior, Peter and Libralato, Mattia and Bennett, David and Shao, Michael and Rhodes, Jason and others},
  journal={JATIS},
  volume={5},
  number={4},
  pages={044005--044005},
  year={2019},
  publisher={Society of Photo-Optical Instrumentation Engineers}
}

@article{Thygesen:inprep,
  title={},
  author={Thygesen, E. and others},
  year={in prep}
}

@article{huston:inprep,
  title={},
  author={Huston, M. and others},
  year={in prep}
}

@article{zohrabi:inprep,
  title={},
  author={Zohrabi, F. and others},
  year={in prep}
}

@ARTICLE{ODonnell:1994a,
       author = {{O'Donnell}, James E.},
        title = "{R v-dependent Optical and Near-Ultraviolet Extinction}",
      journal = {\apj},
         year = 1994,
        month = feb,
       volume = {422},
        pages = {158},
          doi = {10.1086/173713},
       adsurl = {https://ui.adsabs.harvard.edu/abs/1994ApJ...422..158O}
}

@ARTICLE{Cardelli:1989a,
       author = {{Cardelli}, Jason A. and {Clayton}, Geoffrey C. and {Mathis}, John S.},
        title = "{The Relationship between Infrared, Optical, and Ultraviolet Extinction}",
      journal = {\apj},
         year = 1989,
        month = oct,
       volume = {345},
        pages = {245},
          doi = {10.1086/167900},
       adsurl = {https://ui.adsabs.harvard.edu/abs/1989ApJ...345..245C}
}

@article{stassun:2021a,
  title={Parallax systematics and photocenter motions of benchmark eclipsing binaries in Gaia EDR3},
  author={Stassun, Keivan G and Torres, Guillermo},
  journal={The Astrophysical Journal Letters},
  volume={907},
  number={2},
  pages={L33},
  year={2021},
  publisher={IOP Publishing}
}

@article{kluter:2025a,
  title={SynthPop: A New Framework for Synthetic Milky Way Population Generation},
  author={Kl{\"u}ter, Jonas and Huston, Macy J and Aronica, Abigail and Johnson, Samson A and Penny, Matthew T and Newman, Marz and Zohrabi, Farzaneh and Crisp, Alison L and Chevis, Allison},
  journal={\aj},
  volume={169},
  number={6},
  pages={317},
  year={2025},
  publisher={IOP Publishing}
}

@software{huber:2017a,
       author = {{Huber}, Daniel},
        title = "{isoclassify: v1.2}",
         year = 2017,
        month = may,
          eid = {10.5281/zenodo.573372},
          doi = {10.5281/zenodo.573372},
      version = {v1.2},
    publisher = {Zenodo},
       adsurl = {https://ui.adsabs.harvard.edu/abs/2017zndo....573372H}
}

@misc{speagle:2025a,
      title={Deriving Stellar Properties, Distances, and Reddenings using Photometry and Astrometry with BRUTUS}, 
      author={Joshua S. Speagle and Catherine Zucker and Angus Beane and Phillip A. Cargile and Aaron Dotter and Douglas P. Finkbeiner and Gregory M. Green and Benjamin D. Johnson and Edward F. Schlafly and Ana Bonaca and Charlie Conroy and Gwendolyn Eadie and Daniel J. Eisenstein and Alyssa A. Goodman and Jiwon Jesse Han and Harshil M. Kamdar and Rohan Naidu and Hans-Walter Rix and Andrew K. Saydjari and Yuan-Sen Ting and Ioana A. Zelko},
      year={2025},
      eprint={2503.02227},
      archivePrefix={arXiv},
      primaryClass={astro-ph.SR},
      url={https://arxiv.org/abs/2503.02227}, 
}

@article{morton:2015a,
       author = {{Morton}, Timothy D.},
        title = "{isochrones: Stellar model grid package}",
         year = 2015,
        month = mar,
          eid = {ascl:1503.010},
        pages = {ascl:1503.010},
archivePrefix = {ascl},
       eprint = {1503.010},
       adsurl = {https://ui.adsabs.harvard.edu/abs/2015ascl.soft03010M}
}

@article{astropy:2013,
Adsurl = {http://adsabs.harvard.edu/abs/2013A%26A...558A..33A},
Archiveprefix = {arXiv},
Author = {{Astropy Collaboration} and {Robitaille}, T.~P. and {Tollerud}, E.~J. and {Greenfield}, P. and {Droettboom}, M. and {Bray}, E. and {Aldcroft}, T. and {Davis}, M. and {Ginsburg}, A. and {Price-Whelan}, A.~M. and {Kerzendorf}, W.~E. and {Conley}, A. and {Crighton}, N. and {Barbary}, K. and {Muna}, D. and {Ferguson}, H. and {Grollier}, F. and {Parikh}, M.~M. and {Nair}, P.~H. and {Unther}, H.~M. and {Deil}, C. and {Woillez}, J. and {Conseil}, S. and {Kramer}, R. and {Turner}, J.~E.~H. and {Singer}, L. and {Fox}, R. and {Weaver}, B.~A. and {Zabalza}, V. and {Edwards}, Z.~I. and {Azalee Bostroem}, K. and {Burke}, D.~J. and {Casey}, A.~R. and {Crawford}, S.~M. and {Dencheva}, N. and {Ely}, J. and {Jenness}, T. and {Labrie}, K. and {Lim}, P.~L. and {Pierfederici}, F. and {Pontzen}, A. and {Ptak}, A. and {Refsdal}, B. and {Servillat}, M. and {Streicher}, O.},
Doi = {10.1051/0004-6361/201322068},
Eid = {A33},
Eprint = {1307.6212},
Journal = {\aap},
Month = oct,
Pages = {A33},
Primaryclass = {astro-ph.IM},
Title = {{Astropy: A community Python package for astronomy}},
Volume = 558,
Year = 2013}

@ARTICLE{astropy:2018,
       author = {{Astropy Collaboration} and {Price-Whelan}, A.~M. and
         {Sip{\H{o}}cz}, B.~M. and {G{\"u}nther}, H.~M. and {Lim}, P.~L. and
         {Crawford}, S.~M. and {Conseil}, S. and {Shupe}, D.~L. and
         {Craig}, M.~W. and {Dencheva}, N. and {Ginsburg}, A. and {Vand
        erPlas}, J.~T. and {Bradley}, L.~D. and {P{\'e}rez-Su{\'a}rez}, D. and
         {de Val-Borro}, M. and {Aldcroft}, T.~L. and {Cruz}, K.~L. and
         {Robitaille}, T.~P. and {Tollerud}, E.~J. and {Ardelean}, C. and
         {Babej}, T. and {Bach}, Y.~P. and {Bachetti}, M. and {Bakanov}, A.~V. and
         {Bamford}, S.~P. and {Barentsen}, G. and {Barmby}, P. and
         {Baumbach}, A. and {Berry}, K.~L. and {Biscani}, F. and {Boquien}, M. and
         {Bostroem}, K.~A. and {Bouma}, L.~G. and {Brammer}, G.~B. and
         {Bray}, E.~M. and {Breytenbach}, H. and {Buddelmeijer}, H. and
         {Burke}, D.~J. and {Calderone}, G. and {Cano Rodr{\'\i}guez}, J.~L. and
         {Cara}, M. and {Cardoso}, J.~V.~M. and {Cheedella}, S. and {Copin}, Y. and
         {Corrales}, L. and {Crichton}, D. and {D'Avella}, D. and {Deil}, C. and
         {Depagne}, {\'E}. and {Dietrich}, J.~P. and {Donath}, A. and
         {Droettboom}, M. and {Earl}, N. and {Erben}, T. and {Fabbro}, S. and
         {Ferreira}, L.~A. and {Finethy}, T. and {Fox}, R.~T. and
         {Garrison}, L.~H. and {Gibbons}, S.~L.~J. and {Goldstein}, D.~A. and
         {Gommers}, R. and {Greco}, J.~P. and {Greenfield}, P. and
         {Groener}, A.~M. and {Grollier}, F. and {Hagen}, A. and {Hirst}, P. and
         {Homeier}, D. and {Horton}, A.~J. and {Hosseinzadeh}, G. and {Hu}, L. and
         {Hunkeler}, J.~S. and {Ivezi{\'c}}, {\v{Z}}. and {Jain}, A. and
         {Jenness}, T. and {Kanarek}, G. and {Kendrew}, S. and {Kern}, N.~S. and
         {Kerzendorf}, W.~E. and {Khvalko}, A. and {King}, J. and {Kirkby}, D. and
         {Kulkarni}, A.~M. and {Kumar}, A. and {Lee}, A. and {Lenz}, D. and
         {Littlefair}, S.~P. and {Ma}, Z. and {Macleod}, D.~M. and
         {Mastropietro}, M. and {McCully}, C. and {Montagnac}, S. and
         {Morris}, B.~M. and {Mueller}, M. and {Mumford}, S.~J. and {Muna}, D. and
         {Murphy}, N.~A. and {Nelson}, S. and {Nguyen}, G.~H. and
         {Ninan}, J.~P. and {N{\"o}the}, M. and {Ogaz}, S. and {Oh}, S. and
         {Parejko}, J.~K. and {Parley}, N. and {Pascual}, S. and {Patil}, R. and
         {Patil}, A.~A. and {Plunkett}, A.~L. and {Prochaska}, J.~X. and
         {Rastogi}, T. and {Reddy Janga}, V. and {Sabater}, J. and
         {Sakurikar}, P. and {Seifert}, M. and {Sherbert}, L.~E. and
         {Sherwood-Taylor}, H. and {Shih}, A.~Y. and {Sick}, J. and
         {Silbiger}, M.~T. and {Singanamalla}, S. and {Singer}, L.~P. and
         {Sladen}, P.~H. and {Sooley}, K.~A. and {Sornarajah}, S. and
         {Streicher}, O. and {Teuben}, P. and {Thomas}, S.~W. and
         {Tremblay}, G.~R. and {Turner}, J.~E.~H. and {Terr{\'o}n}, V. and
         {van Kerkwijk}, M.~H. and {de la Vega}, A. and {Watkins}, L.~L. and
         {Weaver}, B.~A. and {Whitmore}, J.~B. and {Woillez}, J. and
         {Zabalza}, V. and {Astropy Contributors}},
        title = "{The Astropy Project: Building an Open-science Project and Status of the v2.0 Core Package}",
      journal = {\aj},
         year = 2018,
        month = sep,
       volume = {156},
       number = {3},
          eid = {123},
        pages = {123},
          doi = {10.3847/1538-3881/aabc4f},
archivePrefix = {arXiv},
       eprint = {1801.02634},
 primaryClass = {astro-ph.IM},
       adsurl = {https://ui.adsabs.harvard.edu/abs/2018AJ....156..123A}
}

@ARTICLE{astropy:2022,
       author = {{Astropy Collaboration} and {Price-Whelan}, Adrian M. and {Lim}, Pey Lian and {Earl}, Nicholas and {Starkman}, Nathaniel and {Bradley}, Larry and {Shupe}, David L. and {Patil}, Aarya A. and {Corrales}, Lia and {Brasseur}, C.~E. and {N{"o}the}, Maximilian and {Donath}, Axel and {Tollerud}, Erik and {Morris}, Brett M. and {Ginsburg}, Adam and {Vaher}, Eero and {Weaver}, Benjamin A. and {Tocknell}, James and {Jamieson}, William and {van Kerkwijk}, Marten H. and {Robitaille}, Thomas P. and {Merry}, Bruce and {Bachetti}, Matteo and {G{"u}nther}, H. Moritz and {Aldcroft}, Thomas L. and {Alvarado-Montes}, Jaime A. and {Archibald}, Anne M. and {B{'o}di}, Attila and {Bapat}, Shreyas and {Barentsen}, Geert and {Baz{'a}n}, Juanjo and {Biswas}, Manish and {Boquien}, M{'e}d{'e}ric and {Burke}, D.~J. and {Cara}, Daria and {Cara}, Mihai and {Conroy}, Kyle E. and {Conseil}, Simon and {Craig}, Matthew W. and {Cross}, Robert M. and {Cruz}, Kelle L. and {D'Eugenio}, Francesco and {Dencheva}, Nadia and {Devillepoix}, Hadrien A.~R. and {Dietrich}, J{"o}rg P. and {Eigenbrot}, Arthur Davis and {Erben}, Thomas and {Ferreira}, Leonardo and {Foreman-Mackey}, Daniel and {Fox}, Ryan and {Freij}, Nabil and {Garg}, Suyog and {Geda}, Robel and {Glattly}, Lauren and {Gondhalekar}, Yash and {Gordon}, Karl D. and {Grant}, David and {Greenfield}, Perry and {Groener}, Austen M. and {Guest}, Steve and {Gurovich}, Sebastian and {Handberg}, Rasmus and {Hart}, Akeem and {Hatfield-Dodds}, Zac and {Homeier}, Derek and {Hosseinzadeh}, Griffin and {Jenness}, Tim and {Jones}, Craig K. and {Joseph}, Prajwel and {Kalmbach}, J. Bryce and {Karamehmetoglu}, Emir and {Ka{l}uszy{'n}ski}, Miko{l}aj and {Kelley}, Michael S.~P. and {Kern}, Nicholas and {Kerzendorf}, Wolfgang E. and {Koch}, Eric W. and {Kulumani}, Shankar and {Lee}, Antony and {Ly}, Chun and {Ma}, Zhiyuan and {MacBride}, Conor and {Maljaars}, Jakob M. and {Muna}, Demitri and {Murphy}, N.~A. and {Norman}, Henrik and {O'Steen}, Richard and {Oman}, Kyle A. and {Pacifici}, Camilla and {Pascual}, Sergio and {Pascual-Granado}, J. and {Patil}, Rohit R. and {Perren}, Gabriel I. and {Pickering}, Timothy E. and {Rastogi}, Tanuj and {Roulston}, Benjamin R. and {Ryan}, Daniel F. and {Rykoff}, Eli S. and {Sabater}, Jose and {Sakurikar}, Parikshit and {Salgado}, Jes{'u}s and {Sanghi}, Aniket and {Saunders}, Nicholas and {Savchenko}, Volodymyr and {Schwardt}, Ludwig and {Seifert-Eckert}, Michael and {Shih}, Albert Y. and {Jain}, Anany Shrey and {Shukla}, Gyanendra and {Sick}, Jonathan and {Simpson}, Chris and {Singanamalla}, Sudheesh and {Singer}, Leo P. and {Singhal}, Jaladh and {Sinha}, Manodeep and {Sip{H{o}}cz}, Brigitta M. and {Spitler}, Lee R. and {Stansby}, David and {Streicher}, Ole and {{{S}}umak}, Jani and {Swinbank}, John D. and {Taranu}, Dan S. and {Tewary}, Nikita and {Tremblay}, Grant R. and {Val-Borro}, Miguel de and {Van Kooten}, Samuel J. and {Vasovi{'c}}, Zlatan and {Verma}, Shresth and {de Miranda Cardoso}, Jos{'e} Vin{'i}cius and {Williams}, Peter K.~G. and {Wilson}, Tom J. and {Winkel}, Benjamin and {Wood-Vasey}, W.~M. and {Xue}, Rui and {Yoachim}, Peter and {Zhang}, Chen and {Zonca}, Andrea and {Astropy Project Contributors}},
        title = "{The Astropy Project: Sustaining and Growing a Community-oriented Open-source Project and the Latest Major Release (v5.0) of the Core Package}",
      journal = {\apj},
         year = 2022,
        month = aug,
       volume = {935},
       number = {2},
          eid = {167},
        pages = {167},
          doi = {10.3847/1538-4357/ac7c74},
archivePrefix = {arXiv},
       eprint = {2206.14220},
 primaryClass = {astro-ph.IM},
       adsurl = {https://ui.adsabs.harvard.edu/abs/2022ApJ...935..167A}
}

@Article{harris:2020array,
 title         = {Array programming with {NumPy}},
 author        = {Charles R. Harris and K. Jarrod Millman and St{\'{e}}fan J.
                 van der Walt and Ralf Gommers and Pauli Virtanen and David
                 Cournapeau and Eric Wieser and Julian Taylor and Sebastian
                 Berg and Nathaniel J. Smith and Robert Kern and Matti Picus
                 and Stephan Hoyer and Marten H. van Kerkwijk and Matthew
                 Brett and Allan Haldane and Jaime Fern{\'{a}}ndez del
                 R{\'{i}}o and Mark Wiebe and Pearu Peterson and Pierre
                 G{\'{e}}rard-Marchant and Kevin Sheppard and Tyler Reddy and
                 Warren Weckesser and Hameer Abbasi and Christoph Gohlke and
                 Travis E. Oliphant},
 year          = {2020},
 month         = sep,
 journal       = {Nature},
 volume        = {585},
 number        = {7825},
 pages         = {357--362},
 doi           = {10.1038/s41586-020-2649-2},
 publisher     = {Springer Science and Business Media {LLC}},
 url           = {https://doi.org/10.1038/s41586-020-2649-2}
}

@article{skrutskie:2006,
  title={The two micron all sky survey (2MASS)},
  author={Skrutskie, Michael F and Cutri, RM and Stiening, R and Weinberg, MD and Schneider, S and Carpenter, JM and Beichman, Capps and Capps, R and Chester, T and Elias, J and others},
  journal={The Astronomical Journal},
  volume={131},
  number={2},
  pages={1163},
  year={2006},
  publisher={IOP Publishing}
}

@ARTICLE{kervella:2004a,
       author = {{Kervella}, P. and {Th{\'e}venin}, F. and {Di Folco}, E. and {S{\'e}gransan}, D.},
        title = "{The angular sizes of dwarf stars and subgiants. Surface brightness relations calibrated by interferometry}",
      journal = {\aap},
         year = 2004,
        month = oct,
       volume = {426},
        pages = {297-307},
          doi = {10.1051/0004-6361:20035930},
archivePrefix = {arXiv},
       eprint = {astro-ph/0404180},
 primaryClass = {astro-ph},
       adsurl = {https://ui.adsabs.harvard.edu/abs/2004A&A...426..297K}
}

@ARTICLE{adams:2018a,
       author = {{Adams}, Arthur D. and {Boyajian}, Tabetha S. and {von Braun}, Kaspar},
        title = "{Predicting stellar angular diameters from V, I$_{C}$, H and K photometry}",
      journal = {\mnras},
         year = 2018,
        month = jan,
       volume = {473},
       number = {3},
        pages = {3608-3614},
          doi = {10.1093/mnras/stx2367},
archivePrefix = {arXiv},
       eprint = {1709.03902},
 primaryClass = {astro-ph.SR},
       adsurl = {https://ui.adsabs.harvard.edu/abs/2018MNRAS.473.3608A}
}

@article{nataf:2022prop,
  title={Correcting for the Effects of Interstellar Extinction Toward the Roman Galactic Exoplanet Survey Fields},
  author={Nataf, David Moise and Gennaro, Mario and Benjamin, Robert A and Bennett, David P and Brown, Thomas M and Casagrande, Luca and Paladini, Roberta and Penny, Matthew and Poleski, Radoslaw and Sahu, Kailash C and others},
  journal={HST Proposal},
  pages={17173},
  year={2022}
}

@article{nataf:inprep,
  title={},
  author={Nataf, D. and others},
  journal={},
  year={in prep}
}

@article{nataf:2025prop,
       author = {{Nataf}, David Moise and {Gennaro}, Mario and {Benjamin}, Robert A. and {Bennett}, David P. and {Casagrande}, Luca and {Cassisi}, Santi and {Gaudi}, Bernard Scott and {Jencson}, Jacob and {Paladini}, Roberta and {Penny}, Matthew and {Poleski}, Radoslaw and {Sobeck}, Jennifer and {Stassun}, Keivan G. and {Terry}, Sean and {Zasowski}, Gail},
        journal={HST Proposal},
        pages={17923},
        year={2025}
}

@article{terry:2024prop,
       author = {{Terry}, Sean and {Abrams}, Natasha and {Anderson}, Jay and {Beaulieu}, Jean-Philippe and {Bellini}, Andrea and {Bennett}, David P. and {Bhattacharya}, Aparna and {Brashear}, Jonathan and {Gaudi}, Bernard Scott and {Huston}, Macy and {Ishitani Silva}, Stela and {Johnson}, Samson and {Koshimoto}, Naoki and {Lam}, Casey and {Lu}, Jessica Ryan and {Nataf}, David Moise and {Olmschenk}, Greg and {Penny}, Matthew and {Ranc}, Clement and {Rektsini}, Natalia and {Sumi}, Takahiro and {Suzuki}, Daisuke and {Vandorou}, Aikaterini and {Wilson}, Robert F. and {Zang}, Weicheng and {Zohrabi}, Farzaneh},
        journal={HST Proposal},
        pages={17776},
        year={2024}
}

@ARTICLE{nataf:2016a,
       author = {{Nataf}, David M. and {Gonzalez}, Oscar A. and {Casagrande}, Luca and {Zasowski}, Gail and {Wegg}, Christopher and {Wolf}, Christian and {Kunder}, Andrea and {Alonso-Garcia}, Javier and {Minniti}, Dante and {Rejkuba}, Marina and {Saito}, Roberto K. and {Valenti}, Elena and {Zoccali}, Manuela and {Poleski}, Rados{\l}aw and {Pietrzy{\'n}ski}, Grzegorz and {Skowron}, Jan and {Soszy{\'n}ski}, Igor and {Szyma{\'n}ski}, Micha{\l} K. and {Udalski}, Andrzej and {Ulaczyk}, Krzysztof and {Wyrzykowski}, {\L}ukasz},
        title = "{Interstellar extinction curve variations towards the inner Milky Way: a challenge to observational cosmology}",
      journal = {\mnras},
         year = 2016,
        month = mar,
       volume = {456},
       number = {3},
        pages = {2692-2706},
          doi = {10.1093/mnras/stv2843},
archivePrefix = {arXiv},
       eprint = {1510.01321},
 primaryClass = {astro-ph.SR},
       adsurl = {https://ui.adsabs.harvard.edu/abs/2016MNRAS.456.2692N}
}

@ARTICLE{revnivtsev:2010a,
       author = {{Revnivtsev}, M. and {van den Berg}, M. and {Burenin}, R. and {Grindlay}, J.~E. and {Karasev}, D. and {Forman}, W.},
        title = "{Interstellar extinction and the distribution of stellar populations in the direction of the ultra-deep Chandra Galactic field}",
      journal = {\aap},
         year = 2010,
        month = jun,
       volume = {515},
          eid = {A49},
        pages = {A49},
          doi = {10.1051/0004-6361/200913527},
archivePrefix = {arXiv},
       eprint = {1003.2965},
 primaryClass = {astro-ph.GA},
       adsurl = {https://ui.adsabs.harvard.edu/abs/2010A&A...515A..49R}
}

@ARTICLE{wright:2010a,
       author = {{Wright}, Edward L. and {Eisenhardt}, Peter R.~M. and {Mainzer}, Amy K. and {Ressler}, Michael E. and {Cutri}, Roc M. and {Jarrett}, Thomas and {Kirkpatrick}, J. Davy and {Padgett}, Deborah and {McMillan}, Robert S. and {Skrutskie}, Michael and {Stanford}, S.~A. and {Cohen}, Martin and {Walker}, Russell G. and {Mather}, John C. and {Leisawitz}, David and {Gautier}, III, Thomas N. and {McLean}, Ian and {Benford}, Dominic and {Lonsdale}, Carol J. and {Blain}, Andrew and {Mendez}, Bryan and {Irace}, William R. and {Duval}, Valerie and {Liu}, Fengchuan and {Royer}, Don and {Heinrichsen}, Ingolf and {Howard}, Joan and {Shannon}, Mark and {Kendall}, Martha and {Walsh}, Amy L. and {Larsen}, Mark and {Cardon}, Joel G. and {Schick}, Scott and {Schwalm}, Mark and {Abid}, Mohamed and {Fabinsky}, Beth and {Naes}, Larry and {Tsai}, Chao-Wei},
        title = "{The Wide-field Infrared Survey Explorer (WISE): Mission Description and Initial On-orbit Performance}",
      journal = {\aj},
         year = 2010,
        month = dec,
       volume = {140},
       number = {6},
        pages = {1868-1881},
          doi = {10.1088/0004-6256/140/6/1868},
archivePrefix = {arXiv},
       eprint = {1008.0031},
 primaryClass = {astro-ph.IM},
       adsurl = {https://ui.adsabs.harvard.edu/abs/2010AJ....140.1868W}
}

@ARTICLE{skrutskie:2006a,
       author = {{Skrutskie}, M.~F. and {Cutri}, R.~M. and {Stiening}, R. and {Weinberg}, M.~D. and {Schneider}, S. and {Carpenter}, J.~M. and {Beichman}, C. and {Capps}, R. and {Chester}, T. and {Elias}, J. and {Huchra}, J. and {Liebert}, J. and {Lonsdale}, C. and {Monet}, D.~G. and {Price}, S. and {Seitzer}, P. and {Jarrett}, T. and {Kirkpatrick}, J.~D. and {Gizis}, J.~E. and {Howard}, E. and {Evans}, T. and {Fowler}, J. and {Fullmer}, L. and {Hurt}, R. and {Light}, R. and {Kopan}, E.~L. and {Marsh}, K.~A. and {McCallon}, H.~L. and {Tam}, R. and {Van Dyk}, S. and {Wheelock}, S.},
        title = "{The Two Micron All Sky Survey (2MASS)}",
      journal = {\aj},
         year = 2006,
        month = feb,
       volume = {131},
       number = {2},
        pages = {1163-1183},
          doi = {10.1086/498708},
       adsurl = {https://ui.adsabs.harvard.edu/abs/2006AJ....131.1163S}
}

@ARTICLE{saydjari:2023a,
       author = {{Saydjari}, Andrew K. and {Schlafly}, Edward F. and {Lang}, Dustin and {Meisner}, Aaron M. and {Green}, Gregory M. and {Zucker}, Catherine and {Zelko}, Ioana and {Speagle}, Joshua S. and {Daylan}, Tansu and {Lee}, Albert and {Valdes}, Francisco and {Schlegel}, David and {Finkbeiner}, Douglas P.},
        title = "{The Dark Energy Camera Plane Survey 2 (DECaPS2): More Sky, Less Bias, and Better Uncertainties}",
      journal = {\apjs},
         year = 2023,
        month = feb,
       volume = {264},
       number = {2},
          eid = {28},
        pages = {28},
          doi = {10.3847/1538-4365/aca594},
archivePrefix = {arXiv},
       eprint = {2206.11909},
 primaryClass = {astro-ph.GA},
       adsurl = {https://ui.adsabs.harvard.edu/abs/2023ApJS..264...28S}
}

@ARTICLE{saito:2012a,
       author = {{Saito}, R.~K. and {Hempel}, M. and {Minniti}, D. and {Lucas}, P.~W. and {Rejkuba}, M. and {Toledo}, I. and {Gonzalez}, O.~A. and {Alonso-Garc{\'\i}a}, J. and {Irwin}, M.~J. and {Gonzalez-Solares}, E. and {Hodgkin}, S.~T. and {Lewis}, J.~R. and {Cross}, N. and {Ivanov}, V.~D. and {Kerins}, E. and {Emerson}, J.~P. and {Soto}, M. and {Am{\^o}res}, E.~B. and {Gurovich}, S. and {D{\'e}k{\'a}ny}, I. and {Angeloni}, R. and {Beamin}, J.~C. and {Catelan}, M. and {Padilla}, N. and {Zoccali}, M. and {Pietrukowicz}, P. and {Moni Bidin}, C. and {Mauro}, F. and {Geisler}, D. and {Folkes}, S.~L. and {Sale}, S.~E. and {Borissova}, J. and {Kurtev}, R. and {Ahumada}, A.~V. and {Alonso}, M.~V. and {Adamson}, A. and {Arias}, J.~I. and {Bandyopadhyay}, R.~M. and {Barb{\'a}}, R.~H. and {Barbuy}, B. and {Baume}, G.~L. and {Bedin}, L.~R. and {Bellini}, A. and {Benjamin}, R. and {Bica}, E. and {Bonatto}, C. and {Bronfman}, L. and {Carraro}, G. and {Chen{\`e}}, A.~N. and {Clari{\'a}}, J.~J. and {Clarke}, J.~R.~A. and {Contreras}, C. and {Corvill{\'o}n}, A. and {de Grijs}, R. and {Dias}, B. and {Drew}, J.~E. and {Fari{\~n}a}, C. and {Feinstein}, C. and {Fern{\'a}ndez-Laj{\'u}s}, E. and {Gamen}, R.~C. and {Gieren}, W. and {Goldman}, B. and {Gonz{\'a}lez-Fern{\'a}ndez}, C. and {Grand}, R.~J.~J. and {Gunthardt}, G. and {Hambly}, N.~C. and {Hanson}, M.~M. and {He{\l}miniak}, K.~G. and {Hoare}, M.~G. and {Huckvale}, L. and {Jord{\'a}n}, A. and {Kinemuchi}, K. and {Longmore}, A. and {L{\'o}pez-Corredoira}, M. and {Maccarone}, T. and {Majaess}, D. and {Mart{\'\i}n}, E.~L. and {Masetti}, N. and {Mennickent}, R.~E. and {Mirabel}, I.~F. and {Monaco}, L. and {Morelli}, L. and {Motta}, V. and {Palma}, T. and {Parisi}, M.~C. and {Parker}, Q. and {Pe{\~n}aloza}, F. and {Pietrzy{\'n}ski}, G. and {Pignata}, G. and {Popescu}, B. and {Read}, M.~A. and {Rojas}, A. and {Roman-Lopes}, A. and {Ruiz}, M.~T. and {Saviane}, I. and {Schreiber}, M.~R. and {Schr{\"o}der}, A.~C. and {Sharma}, S. and {Smith}, M.~D. and {Sodr{\'e}}, L. and {Stead}, J. and {Stephens}, A.~W. and {Tamura}, M. and {Tappert}, C. and {Thompson}, M.~A. and {Valenti}, E. and {Vanzi}, L. and {Walton}, N.~A. and {Weidmann}, W. and {Zijlstra}, A.},
        title = "{VVV DR1: The first data release of the Milky Way bulge and southern plane from the near-infrared ESO public survey VISTA variables in the V{\'\i}a L{\'a}ctea}",
      journal = {\aap},
         year = 2012,
        month = jan,
       volume = {537},
          eid = {A107},
        pages = {A107},
          doi = {10.1051/0004-6361/201118407},
archivePrefix = {arXiv},
       eprint = {1111.5511},
 primaryClass = {astro-ph.GA},
       adsurl = {https://ui.adsabs.harvard.edu/abs/2012A&A...537A.107S}
}

@ARTICLE{brown:2011a,
       author = {{Brown}, Timothy M. and {Latham}, David W. and {Everett}, Mark E. and {Esquerdo}, Gilbert A.},
        title = "{Kepler Input Catalog: Photometric Calibration and Stellar Classification}",
      journal = {\aj},
         year = 2011,
        month = oct,
       volume = {142},
       number = {4},
          eid = {112},
        pages = {112},
          doi = {10.1088/0004-6256/142/4/112},
archivePrefix = {arXiv},
       eprint = {1102.0342},
 primaryClass = {astro-ph.SR},
       adsurl = {https://ui.adsabs.harvard.edu/abs/2011AJ....142..112B}
}

@ARTICLE{stassun:2019a,
       author = {{Stassun}, Keivan G. and {Oelkers}, Ryan J. and {Paegert}, Martin and {Torres}, Guillermo and {Pepper}, Joshua and {De Lee}, Nathan and {Collins}, Kevin and {Latham}, David W. and {Muirhead}, Philip S. and {Chittidi}, Jay and {Rojas-Ayala}, B{\'a}rbara and {Fleming}, Scott W. and {Rose}, Mark E. and {Tenenbaum}, Peter and {Ting}, Eric B. and {Kane}, Stephen R. and {Barclay}, Thomas and {Bean}, Jacob L. and {Brassuer}, C.~E. and {Charbonneau}, David and {Ge}, Jian and {Lissauer}, Jack J. and {Mann}, Andrew W. and {McLean}, Brian and {Mullally}, Susan and {Narita}, Norio and {Plavchan}, Peter and {Ricker}, George R. and {Sasselov}, Dimitar and {Seager}, S. and {Sharma}, Sanjib and {Shiao}, Bernie and {Sozzetti}, Alessandro and {Stello}, Dennis and {Vanderspek}, Roland and {Wallace}, Geoff and {Winn}, Joshua N.},
        title = "{The Revised TESS Input Catalog and Candidate Target List}",
      journal = {\aj},
         year = 2019,
        month = oct,
       volume = {158},
       number = {4},
          eid = {138},
        pages = {138},
          doi = {10.3847/1538-3881/ab3467},
archivePrefix = {arXiv},
       eprint = {1905.10694},
 primaryClass = {astro-ph.SR},
       adsurl = {https://ui.adsabs.harvard.edu/abs/2019AJ....158..138S}
}

@ARTICLE{wilson:2023a,
       author = {{Wilson}, Robert F. and {Barclay}, Thomas and {Powell}, Brian P. and {Schlieder}, Joshua and {Hedges}, Christina and {Montet}, Benjamin T. and {Quintana}, Elisa and {Mcdonald}, Iain and {Penny}, Matthew T. and {Espinoza}, N{\'e}stor and {Kerins}, Eamonn},
        title = "{Transiting Exoplanet Yields for the Roman Galactic Bulge Time Domain Survey Predicted from Pixel-level Simulations}",
      journal = {\apjs},
         year = 2023,
        month = nov,
       volume = {269},
       number = {1},
          eid = {5},
        pages = {5},
          doi = {10.3847/1538-4365/acf3df},
archivePrefix = {arXiv},
       eprint = {2305.16204},
 primaryClass = {astro-ph.EP},
       adsurl = {https://ui.adsabs.harvard.edu/abs/2023ApJS..269....5W}
}

@ARTICLE{terry:2020a,
       author = {{Terry}, Sean K. and {Barry}, Richard K. and {Bennett}, David P. and {Bhattacharya}, Aparna and {Anderson}, Jay and {Penny}, Matthew T.},
        title = "{Comparing Observed Stellar Kinematics and Surface Densities in a Low-latitude Bulge Field to Galactic Population Synthesis Models}",
      journal = {\apj},
         year = 2020,
        month = feb,
       volume = {889},
       number = {2},
          eid = {126},
        pages = {126},
          doi = {10.3847/1538-4357/ab629b},
archivePrefix = {arXiv},
       eprint = {1910.02297},
 primaryClass = {astro-ph.SR},
       adsurl = {https://ui.adsabs.harvard.edu/abs/2020ApJ...889..126T}
}

@article{robin:2003a,
  title={A synthetic view on structure and evolution of the Milky Way},
  author={Robin, Annie C and Reyl{\'e}, C and Derri{\`e}re, S and Picaud, S},
  journal={Astronomy \& Astrophysics},
  volume={409},
  number={2},
  pages={523--540},
  year={2003},
  publisher={EDP Sciences}
}

@article{lindegren:2012a,
  title={The astrometric core solution for the gaia mission-overview of models, algorithms, and software implementation},
  author={Lindegren, Lennart and Lammers, Uwe and Hobbs, David and O’Mullane, William and Bastian, Ulrich and Hernandez, Jose},
  journal={Astronomy \& Astrophysics},
  volume={538},
  pages={A78},
  year={2012},
  publisher={EDP Sciences}
}

@article{vallenari:2023a,
  title={Gaia data release 3-summary of the content and survey properties},
  author={Vallenari, Antonella and Brown, Anthony GA and Prusti, Timo and De Bruijne, Jos HJ and Arenou, F and Babusiaux, Carine and Biermann, Michael and Creevey, Orlagh L and Ducourant, Christine and Evans, Dafydd Wyn and others},
  journal={Astronomy \& Astrophysics},
  volume={674},
  pages={A1},
  year={2023},
  publisher={EDP sciences}
}

@article{rektsini:inprep,
  title={},
  author={Rektsini, N. and others},
  year={in prep}
}

@article{terry:inprep,
  title={},
  author={Terry, S. K. and others},
  year={in prep}
}

@article{bhattacharya:2023a,
  title={Confirmation of Color-dependent Centroid Shift Measured After 1.8 Years with HST},
  author={Bhattacharya, Aparna and Bennett, David P and Beaulieu, Jean Philippe and Bond, Ian A and Koshimoto, Naoki and Lu, Jessica R and Blackman, Joshua W and Ranc, Cl{\'e}ment and Vandorou, Aikaterini and Terry, Sean K and others},
  journal={The Astronomical Journal},
  volume={165},
  number={5},
  pages={206},
  year={2023},
  publisher={IOP Publishing}
}

@article{terry:2024a,
  title={Unveiling MOA-2007-BLG-192: An M Dwarf Hosting a Likely Super-Earth},
  author={Terry, Sean K and Beaulieu, Jean-Philippe and Bennett, David P and Hamdorf, Euan and Bhattacharya, Aparna and Chaudhry, Viveka and Cole, Andrew A and Koshimoto, Naoki and Anderson, Jay and Bachelet, Etienne and others},
  journal={The Astronomical Journal},
  volume={168},
  number={2},
  pages={72},
  year={2024},
  publisher={IOP Publishing}
}

@ARTICLE{choi:2016a,
       author = {{Choi}, Jieun and {Dotter}, Aaron and {Conroy}, Charlie and {Cantiello}, Matteo and {Paxton}, Bill and {Johnson}, Benjamin D.},
        title = "{Mesa Isochrones and Stellar Tracks (MIST). I. Solar-scaled Models}",
      journal = {\apj},
         year = 2016,
        month = jun,
       volume = {823},
       number = {2},
          eid = {102},
        pages = {102},
          doi = {10.3847/0004-637X/823/2/102},
archivePrefix = {arXiv},
       eprint = {1604.08592},
 primaryClass = {astro-ph.SR},
       adsurl = {https://ui.adsabs.harvard.edu/abs/2016ApJ...823..102C}
}

@ARTICLE{dotter:2016a,
       author = {{Dotter}, Aaron},
        title = "{MESA Isochrones and Stellar Tracks (MIST) 0: Methods for the Construction of Stellar Isochrones}",
      journal = {\apjs},
         year = 2016,
        month = jan,
       volume = {222},
       number = {1},
          eid = {8},
        pages = {8},
          doi = {10.3847/0067-0049/222/1/8},
archivePrefix = {arXiv},
       eprint = {1601.05144},
 primaryClass = {astro-ph.SR},
       adsurl = {https://ui.adsabs.harvard.edu/abs/2016ApJS..222....8D}
}

@ARTICLE{sumi:2023a,
       author = {{Sumi}, Takahiro and {Koshimoto}, Naoki and {Bennett}, David P. and {Rattenbury}, Nicholas J. and {Abe}, Fumio and {Barry}, Richard and {Bhattacharya}, Aparna and {Bond}, Ian A. and {Fujii}, Hirosane and {Fukui}, Akihiko and {Hamada}, Ryusei and {Hirao}, Yuki and {Silva}, Stela Ishitani and {Itow}, Yoshitaka and {Kirikawa}, Rintaro and {Kondo}, Iona and {Matsubara}, Yutaka and {Miyazaki}, Shota and {Muraki}, Yasushi and {Olmschenk}, Greg and {Ranc}, Cl{\'e}ment and {Satoh}, Yuki and {Suzuki}, Daisuke and {Tomoyoshi}, Mio and {Tristram}, Paul. J. and {Vandorou}, Aikaterini and {Yama}, Hibiki and {Yamashita}, Kansuke},
        title = "{Free-floating Planet Mass Function from MOA-II 9 yr Survey toward the Galactic Bulge}",
      journal = {\aj},
         year = 2023,
        month = sep,
       volume = {166},
       number = {3},
          eid = {108},
        pages = {108},
          doi = {10.3847/1538-3881/ace688},
archivePrefix = {arXiv},
       eprint = {2303.08280},
 primaryClass = {astro-ph.EP},
       adsurl = {https://ui.adsabs.harvard.edu/abs/2023AJ....166..108S}
}

@article{kim:2016a,
  title={KMTNET: a network of 1.6 m wide-field optical telescopes installed at three southern observatories},
  author={Kim, Seung-Lee and Lee, Chung-Uk and Park, Byeong-Gon and Kim, Dong-Jin and Cha, Sang-Mok and Lee, Yongseok and Han, Cheongho and Chun, Moo-Young and Yuk, Insoo},
  journal={Journal of The Korean Astronomical Society, vol. 49, issue 1, pp. 37-44},
  volume={49},
  pages={37--44},
  year={2016}
}

@article{bennett:2024a,
  title={Keck and Hubble Observations Show that MOA-2008-BLG-379Lb is a Super-Jupiter Orbiting an M Dwarf},
  author={Bennett, David P and Bhattacharya, Aparna and Beaulieu, Jean-Philippe and Koshimoto, Naoki and Blackman, Joshua W and Bond, Ian A and Ranc, Cl{\'e}ment and Rektsini, Natalia and Terry, Sean K and Vandorou, Aikaterini and others},
  journal={The Astronomical Journal},
  volume={168},
  number={1},
  pages={15},
  year={2024},
  publisher={IOP Publishing}
}

@article{terry:2023a,
  title={AIROPA IV: Validating point spread function reconstruction on various science cases},
  author={Terry, Sean K and Lu, Jessica R and Turri, Paolo and Ciurlo, Anna and Gautam, Abhimat and Do, Tuan and Fitzgerald, Michael P and Ghez, Andrea and Hosek Jr, Matthew and Witzel, Gunther},
  journal={JATIS},
  volume={9},
  number={1},
  pages={018003--018003},
  month=mar,
  year={2023a},
  publisher={Society of Photo-Optical Instrumentation Engineers}
}

@article{terry:2023b,
      title={The Galactic Center with Roman}, 
      author={Sean K. Terry and Matthew W. Hosek Jr. and Jessica R. Lu and Casey Lam and Natasha Abrams and Arash Bahramian and Richard Barry and Jean-Phillipe Beaulieu and Aparna Bhattacharya and Devin Chu and Anna Ciurlo and Will Clarkson and Tuan Do and Kareem El-Badry and Ryan Felton and Matthew Freeman and Abhimat Gautam and Andrea Ghez and Daniel Huber and Jason Hunt and Macy Huston and Tharindu Jayasinghe and Naoki Koshimoto and Madeline Lucey and Florian Peißker and Anna Pusack and Clément Ranc and Dominick Rowan and Robyn Sanderson and Rainer Schödel and Richard G. Spencer and Rachel Street and Daisuke Suzuki and Aikaterini Vandorou},
      journal={arXiv preprint arXiv:2306.12485},
      month=jun,
      year={2023b},
}

@article{terry:2025a,
  title={Predictions of the Nancy Grace Roman Space Telescope Galactic Exoplanet Survey. IV. Lens Mass and Distance Measurements},
  author={Terry, Sean K and Bachelet, Etienne and Zohrabi, Farzaneh and Verma, Himanshu and Crisp, Alison and Huston, Macy and McGee, Carissma and Penny, Matthew and Abrams, Natasha S and Albrow, Michael D and others},
  journal={The Astronomical Journal},
  volume={171},
  number={4},
  pages={212},
  year={2026},
  publisher={IOP Publishing}
}

@article{anderson:2022a,
  title={One-Pass HST Photometry with hst1pass},
  author={Anderson, Jay},
  journal={Instrument Science Report WFC3},
  volume={5},
  pages={55},
  year={2022}
}

@article{dong:2009a,
  title={Microlensing event MOA-2007-BLG-400: exhuming the buried signature of a cool, Jovian-mass planet},
  author={Dong, Subo and Bond, IA and Gould, A and Koz{\l}owski, Szymon and Miyake, N and Gaudi, BS and Bennett, DP and Abe, F and Gilmore, AC and Fukui, A and others},
  journal={The Astrophysical Journal},
  volume={698},
  number={2},
  pages={1826},
  year={2009},
  publisher={IOP Publishing}
}

@article{koshimoto:2021a,
  title={A Parametric Galactic Model toward the Galactic Bulge Based on Gaia and Microlensing Data},
  author={Koshimoto, Naoki and Baba, Junichi and Bennett, David P},
  journal={The Astrophysical Journal},
  volume={917},
  number={2},
  pages={78},
  year={2021},
  publisher={IOP Publishing}
}

@article{lam:2022a,
  title={An isolated mass gap black hole or neutron star detected with astrometric microlensing},
  author={Lam, Casey Y and Lu, Jessica R and Udalski, Andrzej and Bond, Ian and Bennett, David P and Skowron, Jan and Mroz, Przemek and Poleski, Radek and Sumi, Takahiro and Szymanski, Michal K and others},
  journal={arXiv preprint arXiv:2202.01903},
  year={2022}
}

@article{udalski:2015a,
  title={OGLE-IV: Fourth Phase of the Optical Gravitational Lensing Experiment},
  author={Udalski, A and Szyma{\'n}ski, MK and Szyma{\'n}ski, G},
  journal={Acta Astronomica},
  volume={65},
  number={1},
  pages={1--38},
  year={2015}
}

@article{udalski:1993a,
  title={The optical gravitational lensing experiment. Discovery of the first candidate microlensing event in the direction of the Galactic Bulge},
  author={Udalski, A and Szymanski, M and Kaluzny, J and Kubiak, M and Krzeminski, W and Mateo, M and Preston, GW and Paczynski, B},
  journal={Acta astronomica},
  volume={43},
  pages={289--294},
  year={1993}
}

@article{johnson:2020a,
  title={Predictions of the Nancy Grace Roman Space Telescope Galactic Exoplanet Survey. II. Free-floating Planet Detection Rates},
  author={Johnson, Samson A and Penny, Matthew and Gaudi, B Scott and Kerins, Eamonn and Rattenbury, Nicholas J and Robin, Annie C and Novati, Sebastiano Calchi and Henderson, Calen B},
  journal={\apj},
  volume={160},
  number={3},
  pages={123},
  year={2020},
  publisher={IOP Publishing}
}

@ARTICLE{bhattacharya:2018a,
       author = {{Bhattacharya}, A. and {Beaulieu}, J. -P. and {Bennett}, D.~P. and
         {Anderson}, J. and {Koshimoto}, N. and {Lu}, J.~R. and {Batista}, V. and
         {Blackman}, J.~W. and {Bond}, I.~A. and {Fukui}, A. and
         {Henderson}, C.~B. and {Hirao}, Y. and {Marquette}, J.~B. and
         {Mroz}, P. and {Ranc}, C. and {Udalski}, A.},
        title = "{WFIRST Exoplanet Mass-measurement Method Finds a Planetary Mass of 39 {\ensuremath{\pm}} 8 M $_{{\ensuremath{\oplus}}}$ for OGLE-2012-BLG-0950Lb}",
      journal = {\aj},
         year = 2018,
        month = dec,
       volume = {156},
       number = {6},
          eid = {289},
        pages = {289},
          doi = {10.3847/1538-3881/aaed46},
archivePrefix = {arXiv},
       eprint = {1809.02654},
 primaryClass = {astro-ph.EP},
       adsurl = {https://ui.adsabs.harvard.edu/abs/2018AJ....156..289B}
}

@ARTICLE{bhattacharya:2017a,
       author = {{Bhattacharya}, A. and {Bennett}, D.~P. and {Anderson}, J. and
         {Bond}, I.~A. and {Gould}, A. and {Batista}, V. and {Beaulieu}, J.~P. and
         {Fouqu{\'e}}, P. and {Marquette}, J.~B. and {Pogge}, R.},
        title = "{The Star Blended with the MOA-2008-BLG-310 Source Is Not the Exoplanet Host Star}",
      journal = {\aj},
         year = 2017,
        month = aug,
       volume = {154},
       number = {2},
          eid = {59},
        pages = {59},
          doi = {10.3847/1538-3881/aa7b80},
archivePrefix = {arXiv},
       eprint = {1703.06947},
 primaryClass = {astro-ph.EP},
       adsurl = {https://ui.adsabs.harvard.edu/abs/2017AJ....154...59B}
}

@ARTICLE{bennett:2015a,
       author = {{Bennett}, D.~P. and {Bhattacharya}, A. and {Anderson}, J. and
         {Bond}, I.~A. and {Anderson}, N. and {Barry}, R. and {Batista}, V. and
         {Beaulieu}, J. -P. and {DePoy}, D.~L. and {Dong}, Subo and
         {Gaudi}, B.~S. and {Gilbert}, E. and {Gould}, A. and {Pfeifle}, R. and
         {Pogge}, R.~W. and {Suzuki}, D. and {Terry}, S. and {Udalski}, A.},
        title = "{Confirmation of the Planetary Microlensing Signal and Star and Planet Mass Determinations for Event OGLE-2005-BLG-169}",
      journal = {\apj},
         year = 2015,
        month = aug,
       volume = {808},
       number = {2},
          eid = {169},
        pages = {169},
          doi = {10.1088/0004-637X/808/2/169},
archivePrefix = {arXiv},
       eprint = {1507.08661},
 primaryClass = {astro-ph.EP},
       adsurl = {https://ui.adsabs.harvard.edu/abs/2015ApJ...808..169B}
}

@ARTICLE{boyajian:2014a,
       author = {{Boyajian}, Tabetha S. and {van Belle}, Gerard and {von Braun}, Kaspar},
        title = "{Stellar Diameters and Temperatures. IV. Predicting Stellar Angular Diameters}",
      journal = {\aj},
         year = 2014,
        month = mar,
       volume = {147},
       number = {3},
          eid = {47},
        pages = {47},
          doi = {10.1088/0004-6256/147/3/47},
archivePrefix = {arXiv},
       eprint = {1311.4901},
 primaryClass = {astro-ph.SR},
       adsurl = {https://ui.adsabs.harvard.edu/abs/2014AJ....147...47B}
}

@ARTICLE{nataf:2013a,
       author = {{Nataf}, David M. and {Gould}, Andrew and {Fouqu{\'e}}, Pascal and
         {Gonzalez}, Oscar A. and {Johnson}, Jennifer A. and {Skowron}, Jan and
         {Udalski}, Andrzej and {Szyma{\'n}ski}, Micha{\l} K. and
         {Kubiak}, Marcin and {Pietrzy{\'n}ski}, Grzegorz and
         {Soszy{\'n}ski}, Igor and {Ulaczyk}, Krzysztof and
         {Wyrzykowski}, {\L}ukasz and {Poleski}, Rados{\l}aw},
        title = "{Reddening and Extinction toward the Galactic Bulge from OGLE-III: The Inner Milky Way's R$_{V}$ \raisebox{-0.5ex}\textasciitilde 2.5 Extinction Curve}",
      journal = {\apj},
         year = 2013,
        month = jun,
       volume = {769},
       number = {2},
          eid = {88},
        pages = {88},
          doi = {10.1088/0004-637X/769/2/88},
archivePrefix = {arXiv},
       eprint = {1208.1263},
 primaryClass = {astro-ph.GA},
       adsurl = {https://ui.adsabs.harvard.edu/abs/2013ApJ...769...88N}
}

@ARTICLE{gaudi:2012a,
       author = {{Gaudi}, B. Scott},
        title = "{Microlensing Surveys for Exoplanets}",
      journal = {\araa},
         year = 2012,
        month = sep,
       volume = {50},
        pages = {411-453},
          doi = {10.1146/annurev-astro-081811-125518},
       adsurl = {https://ui.adsabs.harvard.edu/abs/2012ARA&A..50..411G}
}

@ARTICLE{minniti:2010a,
       author = {{Minniti}, D. and {Lucas}, P.~W. and {Emerson}, J.~P. and
         {Saito}, R.~K. and {Hempel}, M. and {Pietrukowicz}, P. and
         {Ahumada}, A.~V. and {Alonso}, M.~V. and {Alonso-Garcia}, J. and
         {Arias}, J.~I. and {Bandyopadhyay}, R.~M. and {Barb{\'a}}, R.~H. and
         {Barbuy}, B. and {Bedin}, L.~R. and {Bica}, E. and {Borissova}, J. and
         {Bronfman}, L. and {Carraro}, G. and {Catelan}, M. and
         {Clari{\'a}}, J.~J. and {Cross}, N. and {de Grijs}, R. and
         {D{\'e}k{\'a}ny}, I. and {Drew}, J.~E. and {Fari{\~n}a}, C. and
         {Feinstein}, C. and {Fern{\'a}ndez Laj{\'u}s}, E. and {Gamen}, R.~C. and
         {Geisler}, D. and {Gieren}, W. and {Goldman}, B. and {Gonzalez}, O.~A. and
         {Gunthardt}, G. and {Gurovich}, S. and {Hambly}, N.~C. and
         {Irwin}, M.~J. and {Ivanov}, V.~D. and {Jord{\'a}n}, A. and
         {Kerins}, E. and {Kinemuchi}, K. and {Kurtev}, R. and
         {L{\'o}pez-Corredoira}, M. and {Maccarone}, T. and {Masetti}, N. and
         {Merlo}, D. and {Messineo}, M. and {Mirabel}, I.~F. and {Monaco}, L. and
         {Morelli}, L. and {Padilla}, N. and {Palma}, T. and {Parisi}, M.~C. and
         {Pignata}, G. and {Rejkuba}, M. and {Roman-Lopes}, A. and
         {Sale}, S.~E. and {Schreiber}, M.~R. and {Schr{\"o}der}, A.~C. and
         {Smith}, M. and {}, L. Sodr{\'e}, Jr. and {Soto}, M. and {Tamura}, M. and
         {Tappert}, C. and {Thompson}, M.~A. and {Toledo}, I. and {Zoccali}, M. and
         {Pietrzynski}, G.},
        title = "{VISTA Variables in the Via Lactea (VVV): The public ESO near-IR variability survey of the Milky Way}",
      journal = {\na},
         year = 2010,
        month = jul,
       volume = {15},
       number = {5},
        pages = {433-443},
          doi = {10.1016/j.newast.2009.12.002},
archivePrefix = {arXiv},
       eprint = {0912.1056},
 primaryClass = {astro-ph.GA},
       adsurl = {https://ui.adsabs.harvard.edu/abs/2010NewA...15..433M}
}

@ARTICLE{bennett:2007a,
       author = {{Bennett}, David P. and {Anderson}, Jay and {Gaudi}, B. Scott},
        title = "{Characterization of Gravitational Microlensing Planetary Host Stars}",
      journal = {\apj},
         year = 2007,
        month = may,
       volume = {660},
       number = {1},
        pages = {781-790},
          doi = {10.1086/513013},
archivePrefix = {arXiv},
       eprint = {astro-ph/0611448},
 primaryClass = {astro-ph},
       adsurl = {https://ui.adsabs.harvard.edu/abs/2007ApJ...660..781B}
}

@ARTICLE{sumi:2003a,
       author = {{Sumi}, T. and {Abe}, F. and {Bond}, I.~A. and {Dodd}, R.~J. and
         {Hearnshaw}, J.~B. and {Honda}, M. and {Honma}, M. and {Kan-ya}, Y. and
         {Kilmartin}, P.~M. and {Masuda}, K. and {Matsubara}, Y. and
         {Muraki}, Y. and {Nakamura}, T. and {Nishi}, R. and {Noda}, S. and
         {Ohnishi}, K. and {Petterson}, O.~K.~L. and {Rattenbury}, N.~J. and
         {Reid}, M. and {Saito}, To. and {Saito}, Y. and {Sato}, H. and
         {Sekiguchi}, M. and {Skuljan}, J. and {Sullivan}, D.~J. and
         {Takeuti}, M. and {Tristram}, P.~J. and {Wilkinson}, S. and
         {Yanagisawa}, T. and {Yock}, P.~C.~M.},
        title = "{Microlensing Optical Depth toward the Galactic Bulge from Microlensing Observations in Astrophysics Group Observations during 2000 with Difference Image Analysis}",
      journal = {\apj},
         year = 2003,
        month = jul,
       volume = {591},
       number = {1},
        pages = {204-227},
          doi = {10.1086/375212},
archivePrefix = {arXiv},
       eprint = {astro-ph/0207604},
 primaryClass = {astro-ph},
       adsurl = {https://ui.adsabs.harvard.edu/abs/2003ApJ...591..204S}
}

@ARTICLE{bond:2001a,
       author = {{Bond}, I.~A. and {Abe}, F. and {Dodd}, R.~J. and {Hearnshaw}, J.~B. and
         {Honda}, M. and {Jugaku}, J. and {Kilmartin}, P.~M. and {Marles}, A. and
         {Masuda}, K. and {Matsubara}, Y. and {Muraki}, Y. and {Nakamura}, T. and
         {Nankivell}, G. and {Noda}, S. and {Noguchi}, C. and {Ohnishi}, K. and
         {Rattenbury}, N.~J. and {Reid}, M. and {Saito}, To. and {Sato}, H. and
         {Sekiguchi}, M. and {Skuljan}, J. and {Sullivan}, D.~J. and {Sumi}, T. and
         {Takeuti}, M. and {Watase}, Y. and {Wilkinson}, S. and {Yamada}, R. and
         {Yanagisawa}, T. and {Yock}, P.~C.~M.},
        title = "{Real-time difference imaging analysis of MOA Galactic bulge observations during 2000}",
      journal = {\mnras},
         year = 2001,
        month = nov,
       volume = {327},
       number = {3},
        pages = {868-880},
          doi = {10.1046/j.1365-8711.2001.04776.x},
archivePrefix = {arXiv},
       eprint = {astro-ph/0102181},
 primaryClass = {astro-ph},
       adsurl = {https://ui.adsabs.harvard.edu/abs/2001MNRAS.327..868B}
}

@ARTICLE{delfosse:2000a,
       author = {{Delfosse}, X. and {Forveille}, T. and {S{\'e}gransan}, D. and {Beuzit}, J. -L. and {Udry}, S. and {Perrier}, C. and {Mayor}, M.},
        title = "{Accurate masses of very low mass stars. IV. Improved mass-luminosity relations}",
      journal = {\aap},
         year = 2000,
        month = dec,
       volume = {364},
        pages = {217-224},
          doi = {10.48550/arXiv.astro-ph/0010586},
archivePrefix = {arXiv},
       eprint = {astro-ph/0010586},
 primaryClass = {astro-ph},
       adsurl = {https://ui.adsabs.harvard.edu/abs/2000A&A...364..217D}
}

@ARTICLE{henry:1993a,
       author = {{Henry}, Todd J. and {McCarthy}, Donald W., Jr.},
        title = "{The Mass-Luminosity Relation for Stars of Mass 1.0 to 0.08M(solar)}",
      journal = {\aj},
         year = 1993,
        month = aug,
       volume = {106},
        pages = {773},
          doi = {10.1086/116685},
       adsurl = {https://ui.adsabs.harvard.edu/abs/1993AJ....106..773H}
}

@ARTICLE{gould:1992a,
       author = {{Gould}, Andrew and {Loeb}, Abraham},
        title = "{Discovering Planetary Systems through Gravitational Microlenses}",
      journal = {\apj},
         year = 1992,
        month = sep,
       volume = {396},
        pages = {104},
          doi = {10.1086/171700},
       adsurl = {https://ui.adsabs.harvard.edu/abs/1992ApJ...396..104G}
}

@ARTICLE{mao:1991a,
       author = {{Mao}, Shude and {Paczynski}, Bohdan},
        title = "{Gravitational Microlensing by Double Stars and Planetary Systems}",
      journal = {\apjl},
         year = 1991,
        month = jun,
       volume = {374},
        pages = {L37},
          doi = {10.1086/186066},
       adsurl = {https://ui.adsabs.harvard.edu/abs/1991ApJ...374L..37M}
}

@ARTICLE{penny19,
       author = {{Penny}, Matthew T. and {Gaudi}, B. Scott and {Kerins}, Eamonn and
         {Rattenbury}, Nicholas J. and {Mao}, Shude and {Robin}, Annie C. and
         {Calchi Novati}, Sebastiano},
        title = "{Predictions of the WFIRST Microlensing Survey. I. Bound Planet Detection Rates}",
      journal = {\apjs},
         year = "2019",
        month = "Mar",
       volume = {241},
       number = {1},
          eid = {3},
        pages = {3},
          doi = {10.3847/1538-4365/aafb69},
archivePrefix = {arXiv},
       eprint = {1808.02490},
 primaryClass = {astro-ph.EP},
       adsurl = {https://ui.adsabs.harvard.edu/abs/2019ApJS..241....3P}
}

@ARTICLE{Surot:2020a,
       author = {{Surot}, F. and {Valenti}, E. and {Gonzalez}, O.~A. and {Zoccali}, M. and {S{\"o}kmen}, E. and {Hidalgo}, S.~L. and {Minniti}, D.},
        title = "{Mapping the stellar age of the Milky Way bulge with the VVV. III. High-resolution reddening map}",
      journal = {\aap},
         year = 2020,
        month = dec,
       volume = {644},
          eid = {A140},
        pages = {A140},
          doi = {10.1051/0004-6361/202038346},
archivePrefix = {arXiv},
       eprint = {2010.02723},
 primaryClass = {astro-ph.GA},
       adsurl = {https://ui.adsabs.harvard.edu/abs/2020A&A...644A.140S}
}

@article{mroz:2019a,
  title={Microlensing optical depth and event rate toward the Galactic bulge from eight years of OGLE-IV observations},
  author={Mroz, P and Udalski, A and Skowron, J and Szymanski, MK and Soszynski, I and Wyrzykowski, L and Pietrukowicz, P and Kozlowski, S and Poleski, R and Ulaczyk, K and others},
  journal={arXiv:1906.02210},
  year={2019}}

@article{simion:2017a,
  title={A parametric description of the 3D structure of the Galactic bar/bulge using the VVV survey},
  author={Simion, Iulia T and Belokurov, Vasily and Irwin, Mike and Koposov, Sergey E and Gonzalez-Fernandez, Carlos and Robin, Annie C and Shen, Juntai and Li, Z-Y},
  journal={MNRAS},
  volume={471},
  number={4},
  pages={4323--4344},
  year={2017},
  publisher={Oxford University Press}}

@article{montet:2017a,
  title={Measuring the galactic distribution of transiting planets with WFIRST},
  author={Montet, Benjamin T and Yee, Jennifer C and Penny, Matthew T},
  journal={Publications of the Astronomical Society of the Pacific},
  volume={129},
  number={974},
  pages={044401},
  year={2017},
  publisher={IOP Publishing}}

@article{kroupa:2001a,
  title={On the variation of the initial mass function},
  author={Kroupa, Pavel},
  journal={MNRAS},
  volume={322},
  number={2},
  pages={231--246},
  year={2001},
  publisher={Blackwell Science Ltd Oxford, UK}}

@article{hunter:2007a,
  title={Matplotlib: A 2D graphics environment},
  author={Hunter, John D},
  journal={Computing in science \& engineering},
  volume={9},
  number={3},
  pages={90},
  year={2007},
  publisher={IEEE Computer Society}}

@article{cao:2013a,
  title={A new photometric model of the Galactic bar using red clump giants},
  author={Cao, Liang and Mao, Shude and Nataf, David and Rattenbury, Nicholas J and Gould, Andrew},
  journal={MNRAS},
  volume={434},
  number={1},
  pages={595--605},
  year={2013},
  publisher={The Royal Astronomical Society}}

@article{anderson:2008a,
  title={The ACS Survey of globular clusters. V. Generating a comprehensive star catalog for each cluster},
  author={Anderson, Jay and Sarajedini, Ata and Bedin, Luigi R and King, Ivan R and Piotto, Giampaolo and Reid, I Neill and Siegel, Michael and Majewski, Steven R and Paust, Nathaniel EQ and Aparicio, Antonio and others},
  journal={AJ},
  volume={135},
  number={6},
  pages={2055},
  year={2008},
  publisher={IOP Publishing}}

@article{ortolani:1995a,
  title={Near-coeval formation of the Galactic bulge and halo inferred from globular cluster ages},
  author={Ortolani, Sergio and Renzini, Alvio and Gilmozzi, Roberto and Marconi, Gianni and Barbuy, Beatriz and Bica, Eduardo and Rich, R Michael},
  journal={Nature},
  volume={377},
  number={6551},
  pages={701},
  year={1995},
  publisher={Nature Publishing Group}}

@article{brown:2009a,
  title={The WFC3 galactic bulge treasury program: A first look at resolved stellar population tools},
  author={Brown, Thomas M and Sahu, Kailash and Zoccali, Manuela and Renzini, Alvio and Ferguson, Henry C and Anderson, Jay and Smith, Ed and Bond, Howard E and Minniti, Dante and Valenti, Jeff A and others},
  journal={AJ},
  volume={137},
  number={2},
  pages={3172},
  year={2009},
  publisher={IOP Publishing}}

@article{sharma:2011a,
  title={Galaxia: A code to generate a synthetic survey of the Milky Way},
  author={Sharma, Sanjib and Bland-Hawthorn, Joss and Johnston, Kathryn V and Binney, James},
  journal={ApJ},
  volume={730},
  number={1},
  pages={3},
  year={2011},
  publisher={IOP Publishing}}

@article{penny:2019a,
  title={Predictions of the WFIRST Microlensing Survey. I. Bound Planet Detection Rates},
  author={Penny, Matthew T and Gaudi, B Scott and Kerins, Eamonn and Rattenbury, Nicholas J and Mao, Shude and Robin, Annie C and Novati, Sebastiano Calchi},
  journal={ApJS},
  volume={241},
  number={1},
  pages={3},
  year={2019},
  publisher={IOP Publishing}}

@article{wegg:2013a,
  title={Mapping the three-dimensional density of the Galactic bulge with VVV red clump stars},
  author={Wegg, Christopher and Gerhard, Ortwin},
  journal={MNRAS},
  volume={435},
  number={3},
  pages={1874--1887},
  year={2013},
  publisher={Oxford University Press}}

@article{nataf:2010a,
  title={The split red clump of the Galactic Bulge from OGLE-III},
  author={Nataf, DM and Udalski, A and Gould, A and Fouqu{\'e}, P and Stanek, KZ},
  journal={ApJL},
  volume={721},
  number={1},
  pages={L28},
  year={2010},
  publisher={IOP Publishing}}

@article{stanek:1997a,
  title={Modeling the galactic bar using red clump giants},
  author={Stanek, KZ and Udalski, A and Szyma{\'n}ski, M and Ka{\l}u{\.z}ny, J and Kubiak, Z M and Mateo, M and Krzemi{\'n}ski, W},
  journal={ApJ},
  volume={477},
  number={1},
  pages={163},
  year={1997},
  publisher={IOP Publishing}}

@article{stanek:1994a,
  title={Color-Magnitude Diagram Distribution of the Bulge Red Clump Stars-Evidence for the Galactic Bar},
  author={Stanek, KZ and Mateo, M and Udalski, A and Szymanski, M and Kaluzny, J and Kubiak, M},
  journal={ApJ},
  volume={429},
  number={2},
  pages={L73},
  year={1994}}

@article{bernard:2018a,
  title={Star formation history of the Galactic bulge from deep HST imaging of low reddening windows},
  author={Bernard, Edouard J and Schultheis, Mathias and Di Matteo, Paola and Hill, Vanessa and Haywood, Misha and Calamida, Annalisa},
  journal={MNRAS},
  volume={477},
  number={3},
  pages={3507--3519},
  year={2018},
  publisher={Oxford University Press}}

@article{zoccali:2000a,
  title={The Initial Mass Function of the Galactic Bulge down to\~{} 0.15 M☉},
  author={Zoccali, Manuela and Cassisi, Santi and Frogel, Jay A and Gould, Andrew and Ortolani, Sergio and Renzini, Alvio and Rich, R Michael and Stephens, Andrew W},
  journal={ApJ},
  volume={530},
  number={1},
  pages={418},
  year={2000},
  publisher={IOP Publishing}}

@article{gonzalez:2012a,
  title={Reddening and metallicity maps of the Milky Way bulge from VVV and 2MASS-II. The complete high resolution extinction map and implications for Galactic bulge studies},
  author={Gonzalez, OA and Rejkuba, M and Zoccali, M and Valenti, E and Minniti, D and Schultheis, M and Tobar, R and Chen, B},
  journal={A\&A},
  volume={543},
  pages={A13},
  year={2012},
  publisher={EDP Sciences}}

@article{Spergel:2015a,
	Adsurl = {http://adsabs.harvard.edu/abs/2015arXiv150303757S},
	Archiveprefix = {arXiv},
	Author = {{Spergel}, D. and {Gehrels}, N. and {Baltay}, C. and {Bennett}, D. and {Breckinridge}, J. and {Donahue}, M. and {Dressler}, A. and {Gaudi}, B.~S. and {Greene}, T. and {Guyon}, O. and {Hirata}, C. and {Kalirai}, J. and {Kasdin}, N.~J. and {Macintosh}, B. and {Moos}, W. and {Perlmutter}, S. and {Postman}, M. and {Rauscher}, B. and {Rhodes}, J. and {Wang}, Y. and {Weinberg}, D. and {Benford}, D. and {Hudson}, M. and {Jeong}, W.-S. and {Mellier}, Y. and {Traub}, W. and {Yamada}, T. and {Capak}, P. and {Colbert}, J. and {Masters}, D. and {Penny}, M. and {Savransky}, D. and {Stern}, D. and {Zimmerman}, N. and {Barry}, R. and {Bartusek}, L. and {Carpenter}, K. and {Cheng}, E. and {Content}, D. and {Dekens}, F. and {Demers}, R. and {Grady}, K. and {Jackson}, C. and {Kuan}, G. and {Kruk}, J. and {Melton}, M. and {Nemati}, B. and {Parvin}, B. and {Poberezhskiy}, I. and {Peddie}, C. and {Ruffa}, J. and {Wallace}, J.~K. and {Whipple}, A. and {Wollack}, E. and {Zhao}, F.},
	Eprint = {1503.03757},
	Journal = {ArXiv e-prints},
	Month = mar,
	Primaryclass = {astro-ph.IM},
	Title = {{Wide-Field InfrarRed Survey Telescope-Astrophysics Focused Telescope Assets WFIRST-AFTA 2015 Report}},
	Year = 2015}

@article{Rich:2007a,
	Adsurl = {http://adsabs.harvard.edu/abs/2007ApJ...658L..29R},
	Author = {{Rich}, R.~M. and {Reitzel}, D.~B. and {Howard}, C.~D. and {Zhao}, H.},
	Doi = {10.1086/513509},
	Eprint = {astro-ph/0611403},
	Journal = {\apjl},
	Month = mar,
	Pages = {L29-L32},
	Title = {{The Bulge Radial Velocity Assay: Techniques and a Rotation Curve}},
	Volume = 658,
	Year = 2007}

@article{Milone:2012a,
	Adsurl = {http://adsabs.harvard.edu/abs/2012A%26A...540A..16M},
	Archiveprefix = {arXiv},
	Author = {{Milone}, A.~P. and {Piotto}, G. and {Bedin}, L.~R. and {Aparicio}, A. and {Anderson}, J. and {Sarajedini}, A. and {Marino}, A.~F. and {Moretti}, A. and {Davies}, M.~B. and {Chaboyer}, B. and {Dotter}, A. and {Hempel}, M. and {Mar{\'{\i}}n-Franch}, A. and {Majewski}, S. and {Paust}, N.~E.~Q. and {Reid}, I.~N. and {Rosenberg}, A. and {Siegel}, M.},
	Doi = {10.1051/0004-6361/201016384},
	Eid = {A16},
	Eprint = {1111.0552},
	Journal = {\aap},
	Month = apr,
	Pages = {A16},
	Primaryclass = {astro-ph.SR},
	Title = {{The ACS survey of Galactic globular clusters. XII. Photometric binaries along the main sequence}},
	Volume = 540,
	Year = 2012}

@article{Lagioia:2014a,
	Adsurl = {http://adsabs.harvard.edu/abs/2014ApJ...782...50L},
	Archiveprefix = {arXiv},
	Author = {{Lagioia}, E.~P. and {Milone}, A.~P. and {Stetson}, P.~B. and {Bono}, G. and {Prada Moroni}, P.~G. and {Dall'Ora}, M. and {Aparicio}, A. and {Buonanno}, R. and {Calamida}, A. and {Ferraro}, I. and {Gilmozzi}, R. and {Iannicola}, G. and {Matsunaga}, N. and {Monelli}, M. and {Walker}, A.},
	Doi = {10.1088/0004-637X/782/1/50},
	Eid = {50},
	Eprint = {1312.2272},
	Journal = {\apj},
	Month = feb,
	Pages = {50},
	Primaryclass = {astro-ph.SR},
	Title = {{On the Kinematic Separation of Field and Cluster Stars across the Bulge Globular NGC 6528}},
	Volume = 782,
	Year = 2014}

@article{Kozlowski:2006a,
	Adsurl = {http://adsabs.harvard.edu/abs/2006MNRAS.370..435K},
	Author = {{Koz{\l}owski}, S. and {Wo{\'z}niak}, P.~R. and {Mao}, S. and {Smith}, M.~C. and {Sumi}, T. and {Vestrand}, W.~T. and {Wyrzykowski}, {\L}.},
	Doi = {10.1111/j.1365-2966.2006.10487.x},
	Eprint = {astro-ph/0604550},
	Journal = {\mnras},
	Month = jul,
	Pages = {435-443},
	Title = {{Mapping stellar kinematics across the Galactic bar: HST measurements of proper motions in 35 fields}},
	Volume = 370,
	Year = 2006}

@article{Holtzman:1998a,
	Adsurl = {http://adsabs.harvard.edu/abs/1998AJ....115.1946H},
	Author = {{Holtzman}, J.~A. and {Watson}, A.~M. and {Baum}, W.~A. and {Grillmair}, C.~J. and {Groth}, E.~J. and {Light}, R.~M. and {Lynds}, R. and {O'Neil}, Jr., E.~J.},
	Doi = {10.1086/300336},
	Eprint = {astro-ph/9801321},
	Journal = {\aj},
	Month = may,
	Pages = {1946-1957},
	Title = {{The Luminosity Function and Initial Mass Function in the Galactic Bulge}},
	Volume = 115,
	Year = 1998}

@inproceedings{Gennaro:2015a,
	Adsurl = {http://adsabs.harvard.edu/abs/2015ASPC..491..182G},
	Author = {{Gennaro}, M. and {Brown}, T. and {Anderson}, J. and {Avila}, R. and {VandenBerg}, D. and {Sahu}, K. and {Bond}, H. and {Casertano}, S. and {Ferguson}, H. and {Livio}, M. and {Minniti}, D. and {Panagia}, N. and {Renzini}, A. and {Tumlinson}, J. and {Valenti}, E. and {Valenti}, J. and {Zoccali}, M.},
	Booktitle = {Fifty Years of Wide Field Studies in the Southern Hemisphere: Resolved Stellar Populations of the Galactic Bulge and Magellanic Clouds},
	Editor = {},
	Month = may,
	Pages = {182},
	Series = {Astronomical Society of the Pacific Conference Series},
	Title = {{The Initial Mass Function and Star Formation History of the Galactic Bulge from HST*}},
	Volume = 491,
	Year = 2015}

@article{Clarkson:2008a,
	Adsurl = {http://adsabs.harvard.edu/abs/2008ApJ...684.1110C},
	Archiveprefix = {arXiv},
	Author = {{Clarkson}, W. and {Sahu}, K. and {Anderson}, J. and {Smith}, T.~E. and {Brown}, T.~M. and {Rich}, R.~M. and {Casertano}, S. and {Bond}, H.~E. and {Livio}, M. and {Minniti}, D. and {Panagia}, N. and {Renzini}, A. and {Valenti}, J. and {Zoccali}, M.},
	Doi = {10.1086/590378},
	Eprint = {0809.1682},
	Journal = {\apj},
	Month = sep,
	Pages = {1110-1142},
	Title = {{Stellar Proper Motions in the Galactic Bulge from Deep Hubble Space Telescope ACS WFC Photometry}},
	Volume = 684,
	Year = 2008}

@article{Baldwin:2016a,
	Adsurl = {http://adsabs.harvard.edu/abs/2016ApJ...827...12B},
	Archiveprefix = {arXiv},
	Author = {{Baldwin}, A.~T. and {Watkins}, L.~L. and {van der Marel}, R.~P. and {Bianchini}, P. and {Bellini}, A. and {Anderson}, J.},
	Doi = {10.3847/0004-637X/827/1/12},
	Eid = {12},
	Eprint = {1606.00836},
	Journal = {\apj},
	Month = aug,
	Pages = {12},
	Primaryclass = {astro-ph.SR},
	Title = {{Hubble Space Telescope Proper Motion (HSTPROMO) Catalogs of Galactic Globular Clusters. IV. Kinematic Profiles and Average Masses of Blue Straggler Stars}},
	Volume = 827,
	Year = 2016}

\end{document}